\documentclass[prb, aps, twocolumn, asmath, amssymb, floatfix, superscriptaddress,longbibliography]{revtex4-2}
\usepackage[dvips]{graphics}
\usepackage{adjustbox}
\usepackage{color}
\usepackage{float}
\usepackage{natbib}
\usepackage[dvipsnames]{xcolor}
\setcitestyle{numbers}
\setcitestyle{square}
\definecolor{dred}{rgb}{0,0,0.6}
\definecolor{NavyBlue}{rgb}{0, 0, 128}
\definecolor{RoyalBlue}{rgb}{0.255,0.41,0.884}
\usepackage{graphicx}
\usepackage{bm}
\usepackage{amssymb}
\usepackage{amsmath}
\usepackage{mathtools}
\usepackage{dcolumn}   
\usepackage[colorlinks, linkcolor=NavyBlue,citecolor=NavyBlue,urlcolor=NavyBlue]{hyperref}
\usepackage[all]{hypcap} 
\usepackage{physics}
\usepackage{enumerate}
\usepackage{braket}        
\usepackage{overpic}
\usepackage{amsfonts}
\usepackage{dsfont}
\usepackage[mathscr]{eucal}
\usepackage{hyperref}
\usepackage{setspace}
\usepackage{url}

\newcommand{\dg}{\dagger}
\newcommand{\mb}{\mathbf}

\newcommand{\prbsubref}[2]{\hyperref[#1]{\ref*{#1}{(\subref*{#2})}}}

\usepackage[normalem]{ulem}
\definecolor{lime}{HTML}{A6CE39}
\usepackage{sidecap,tikz}
\DeclareRobustCommand{\orcidicon}{\hspace{-1.0mm}
	\begin{tikzpicture}
		\draw[lime, fill=lime] (0.0,0.0) 
		circle [radius=0.15] 
		node[white] {{\fontfamily{qag}\selectfont \tiny \,ID}};
		\draw[white, fill=white] (-0.0525,0.095) 
		circle [radius=0.007];
	\end{tikzpicture}
	\hspace{-3.0mm}
}
\foreach \x in {A, ..., Z}{\expandafter\xdef\csname orcid\x\endcsname{\noexpand\href{https://orcid.org/\csname orcidauthor\x\endcsname}
		{\noexpand\orcidicon}}
}

\begin{document}
\newcommand{\Z}{\mathbb{Z}}

\title{Topological characterization of magnon-polaron bands and thermal Hall conductivity in a frustrated kagome antiferromagnet}

\author{Shreya Debnath$^\S$\orcidA{}}
\email[]{d.shreya@iitg.ac.in} 

\author{Kuntal Bhattacharyya$^\S$\orcidB{}}
\email[]{kuntalphy@iitg.ac.in}

\author{Saurabh Basu}
\email[]{saurabh@iitg.ac.in}
\affiliation{Department of Physics, Indian Institute of Technology Guwahati, Guwahati-781039, Assam, India}

\begin{abstract}
Spin-phonon coupling and its efficacy in inducing multiple topological phase transitions in a frustrated kagome antiferromagnet have been rare in literature. To this end, we study the ramifications of invoking optical phonons in such a system via two different coupling mechanisms, namely, a local and a non-local one, which are distinct in their microscopic origin.
In case of the local spin-phonon coupling, a single phonon mode affects the magnetic interactions, whereas in the non-local case, two neighbouring phonon modes are involved in the energy renormalization, and it would be worthwhile to compare and contrast between the two. To tackle these phonons, we propose an analytic approach involving a canonical spin-Peierls transformation applied to magnons. The formalism renders a hybridization between the magnons and the phonon modes, yielding magnon-polaron quasiparticles. In both the coupling regimes, validations for the topological signatures are systematically derived from the bulk and edge spectral properties of the magnon-polaron bands that are characterized by their corresponding Chern numbers. Thereafter, we investigate transitions from one topological phase to another solely via tuning the spin-phonon coupling strength. Moreover, these transitions significantly impact the behavior of the thermal Hall conductivity that aids in discerning distinct topological phases.
Additionally, the explicit dependencies on the temperature and the external magnetic field are explored in inducing topological phase transitions associated with the magnon-polaron bands.
Thus, our work serves as an ideal platform to probe the interplay of frustrated magnetism and polaronic physics.
\end{abstract}
   
\maketitle


\section{INTRODUCTION}
\label{Introduction}
As an electronic analogue, investigating topological properties in magnetically ordered systems has become an increasingly active area of modern condensed matter research~\cite{McClarty2022,Wang2021,Li2021}. In particular, the disordered spin state, called a quantum spin liquid (QSL), and frustrated magnetism hold significant interest~\cite{Anderson1973,Lhuillier2005,Moessner2006,Balents2010,Zhou2017, Savary2017,Norman2016,Broholm2020}. 
Studies have delineated the effects of quantized spin waves or magnons on the band topology~\cite{McClarty2022,Wang2021,Li2021,Matsumoto2011,MatsumotoII2011,Murakami2017,Katsura2010,Onose2010,Ideue2012,Hirschberger2015,Malki2017,Owerre2017,Laurell2018,Akazawa2020,Dias2023,He2024,Debnath2024,Debnath2025,Lee2015,HirschbergerII2015} in bosonic systems, where magnons, being charge-neutral quasiparticles, are unharmed by the electronic correlations, and thus they avoid Joule heating. Moreover, novel spintronic systems favor strong spin-orbit coupling giving rise to the Dzyaloshinskii-Moriya interaction (DMI)~\cite{dzyaloshinsky1958, moriya1960} that opens up alternative avenues towards studying the nontrivial topology of the magnon bands in both magnetically ordered~\cite{Katsura2010,Onose2010,Ideue2012,Hirschberger2015,Malki2017,Owerre2017,Laurell2018,Akazawa2020,Dias2023,He2024,Debnath2024,Debnath2025} and QSL~\cite{Katsura2010,Lee2015,HirschbergerII2015} systems. In the presence of a finite temperature gradient, the DMI induces a nontrivial Berry curvature (acting as an effective magnetic field) that leads to the magnon thermal Hall effect (MTHE)~\cite{Matsumoto2011,MatsumotoII2011,Murakami2017,Katsura2010,Onose2010,Ideue2012,Hirschberger2015,Malki2017,Owerre2017,Laurell2018,Akazawa2020,Dias2023,He2024,Debnath2024,Debnath2025,Lee2015,HirschbergerII2015}, exploration of which, being active more than ever, is useful for extracting the topological properties of the \textit{magnonics} (magnon spintronics) systems. Materials with such properties are suitable for exploiting futuristic magnonics applications~\cite{Chumak2015,Wang2018,Baltz2018}.  

Of late, geometrically frustrated kagome antiferromagnets (KAFM) have been proven to be an important candidate to host exotic QSL features~\cite{Sachdev1992,Yan2011,Lu2011,Depenbrock2012,Jiang2012,Iqbal2013,Nishimoto2013,Mei2017, Liao2017,He2017,Jiang2019}. Characterized by various elementary excitations, the theoretical studies have speculated on ground state QSL markers of the kagome Heisenberg antiferromagnet, such as the $\mathbb{Z}_2$~\cite{Wang2006,Lu2011,Depenbrock2012,Mei2017,Block2020}, Dirac~\cite{Iqbal2013,Liao2017,He2017,Jiang2019}, topologically entangled~\cite{Jiang2012}, and chiral~\cite{Messio2017,He2024} spin liquid states, and more recently, spinon pair-density wave state~\cite{Duric2025}. Although, among the QSLs chiral spin liquids break the time-reversal symmetry (TRS) and result in formation of Chern bands, it is unclear whether the QSL ground state of the KAFM is gapped~\cite{Yan2011,Lu2011,Depenbrock2012,Mei2017, Depenbrock2012,Iqbal2013,He2017}. 
In addition to these fundamental issues, a KAFM harbors worthwhile quantum signatures induced by DMI~\cite{Elhajal2002,Matan2006,Cepas2008,Huh2010,Zorko2013, Mishchenko2014, Owerre2017,Laurell2018,Akazawa2020,Dias2023,He2024}. The DMI can intrinsically induce long-range order (LRO), commonly termed $\bm {Q}=0$ magnetic order in frustrated kagome magnets, which is otherwise inaccessible at low temperatures. Hence, up to a critical order, DMI suppresses the QSL phase of the KAFM~\cite{Cepas2008}. Past studies have evidenced the LRO in coplanar or noncollinear frustrated KAFM at specific regimes of temperature~\cite{Lee1997,Inami2001,Elhajal2002,Grohol2005,Matan2006,Cepas2008}. However, theoretical predictions on non-coplanar KAFM structures, such as pyrochlore thin film~\cite{Laurell2017}, jarosites~\cite{Wills2001,Nishiyama2003,Yildirim2006,Inami2001,Elhajal2002,Matan2006}, namely $\text{KFe}_3\text{(OH)}_6(\text{SO}_4)_2$, $\text{AgFe}_3\text{(OH)}_6(\text{SO}_4)_2$, etc., Vesignieite  $\text{BaCu}_3\text{V}_2\text{O}_8\text{(OH)}_2$~\cite{Zorko2013}, herbertsmithite $\text{ZnCu}_3\text{(OH)}_6\text{Cl}_2$~\cite{Zorko2008}, and noncollinear canted configurations induced by a magnetic field~\cite{Jeschke2015,Verrier2020,Sheng2025} have been probed to discern MTHE experimentally. Due to the inherent noncoplanarity in several of these KAFMs, a DMI-induced scalar spin chirality, $\chi=\bm S_i.(\bm S_j\times \bm S_k)$ originates which instills a nontrivial magnon topology~\cite{Elhajal2002,Matan2006,Cepas2008,Mishchenko2014,Owerre2017,Laurell2018} and MTHE~\cite{Owerre2017,Laurell2018,Akazawa2020,Dias2023,He2024}. While a field-induced non-coplanar spin chirality is solely capable of manifesting MTHE in KAFMs, with no mandatory requirement of DMI~\cite{Owerre2017}. In contrast, with no magnetic field, the in-plane DMI has been reported to stabilize a canted KAFM configuration with a finite spin chirality that significantly modifies the magnon thermal Hall conductivity~\cite{Laurell2018} corresponding to each of the Chern (magnon) bands. Moreover, KAFMs host a \text{``spinon quantum spin Hall effect''} in the time-reversal symmetric gapped QSL phase and facilitate a continuous phase transition into this phase with the inclusion of an extended nearest neighbour hopping that \textit{melts} the DMI-induced $\bm Q=0$ ordering~\cite{He2024}. Hence, this gapped QSL phase with oppositely moving spinon chiral edges (possessing spin Chern numbers, $C_{\uparrow,\downarrow}=\pm 1$) renders itself as a magnetic analogue of the $\mathbb{Z}_2$ topological insulator. However, even with all of these proposals being accounted for, there is still enough scope to explore quasiparticle excitonic topology and MTHE of the KAFM magnon bands.

Intriguingly, other variants of quasiparticles, such as phonons (associated with the lattice vibrations), also corroborate nontrivial topological phenomena. Akin to phonon-induced band topology in electronic systems~\cite{Garate2013,Li2013,Cangemi2019,Hu2021,Medina2022,Lu2023,Islam2024,Bhattacharyya2024}, these vibrational modes can hybridize with magnons, resulting in magnon-phonon (also termed \textit{magnon-polaron}) quasiparticle coupled topological signatures in magnetically ordered systems~\cite{Takahashi2016,Park2016,Park2019,Zhang2019,Go2019, Zhang2020,Bao2020,Sheikhi2021,Liu2021,Mai2021,Chen2021,Li2022,Ma2022,Tiwari2023,Lyons2023,Klogetvedt2023,Li2023,Mella2024,Metzger2024,Wu2025}. To mention a few, the hybridization effects of the magnon and phonon bands are mostly explored in square~\cite{Zhang2019, Zhang2020,Liu2021}, honeycomb~\cite{Sheikhi2021,Ma2022,Tiwari2023,Klogetvedt2023,Li2023,Mella2024} lattices, and two-dimensional (2D) antiferromagnets~\cite{Park2016,Park2019,Zhang2020,Liu2021,Ma2022,Lyons2023,Li2023}. It has been claimed that hybridization of these quasiparticles can occur with (via breaking the mirror symmetry)~\cite{Bao2020,Zhang2019,Sheikhi2021, Ma2022, Klogetvedt2023} or without~\cite{Go2019,Park2019,Takahashi2016} DMI or through dipolar interactions~\cite{Takahashi2016}, which results in anticrossing of these excitonic bands accompanied by a nontrivial Berry curvature that induces magnon-polaron thermal Hall effect. The hybridization in the absence of any DMI is usually explained by magnetostriction or magnetoelastic wave~\cite{Kittel1958,Go2019,Park2019}, which forms a magnon-polaron state mainly through the acoustic phonon modes. While the role of optical phonon modes has also been examined in manipulating the Chern numbers of the hybridized magnon-polaron bands~\cite{Chen2021,Ma2022}, such studies are non-existent in the context of a kagome lattice. Moreover, magnon-phonon topological physics in the kagome structure is not well established in the literature, especially on a KAFM configuration. Hence, a closer inspection of the magnon-\textit{optical} phonon coupled quasiparticle scenario (in terms of magnon-polaron bound state) on a KAFM may unravel important topological properties of frustrated magnetism, which, to the best of our knowledge, has not been reported yet.

\begin{figure}[t]
    \centering
    \includegraphics[width=1\columnwidth]{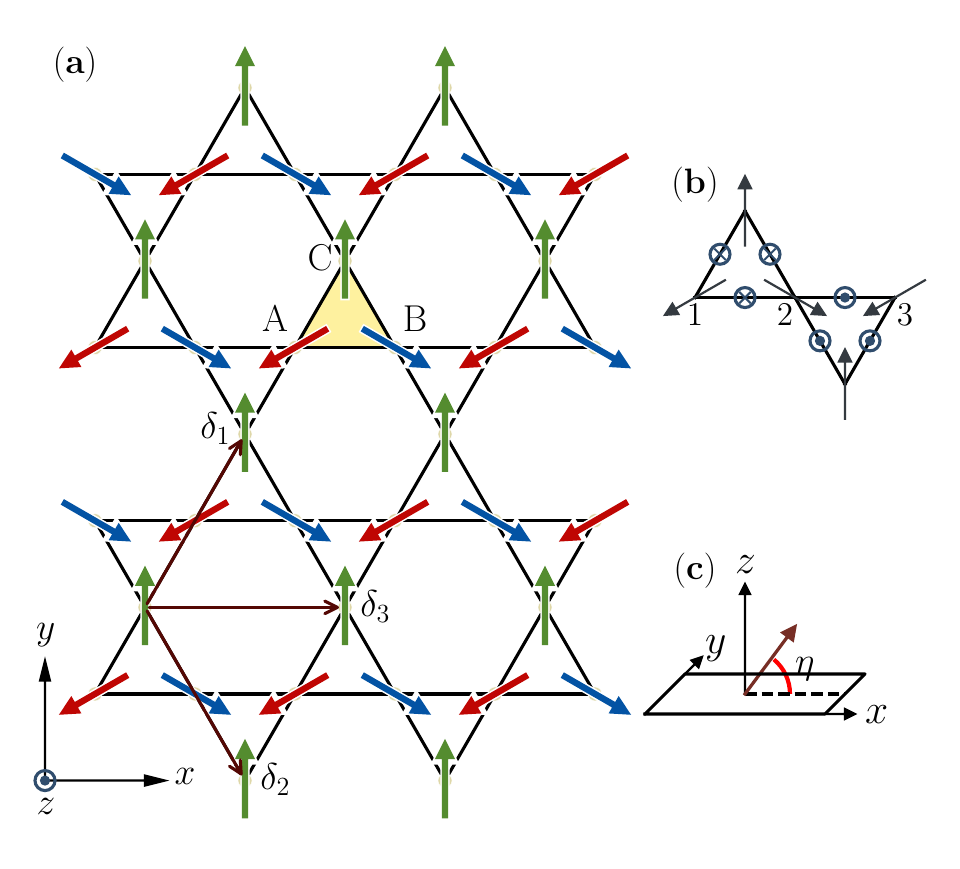}
    \caption{Schematic diagram of a kagome antiferromagnet is displayed in (a), where red, blue, and green arrows refer to the spin orientation of A, B, and C sublattices, respectively, and $\mathbf{\delta}_1$, $\mathbf{\delta}_2$ and $\mb{\delta}_3$ denote three lattice vectors. (b) The direction of the out-of-plane DMI is shown, which acts downward from site 1 to 2, and upward from site 2 to 3, while (c) illustrates that in the presence of an external magnetic field, the spins become canted with an angle $\eta$ with respect to the $x$-$y$ plane.}
    \label{Kagome}
\end{figure}

In this study, we explore the effects of magnon-polaron excitations on a 2D non-coplanar KAFM (as schematized in Fig.~\ref{Kagome}(a)) in the presence of an out-of-plane DMI ($D$) and an external magnetic field ($B_0$). Additionally, we formulate our magnon model with the nearest neighbour (NN) and the next-nearest neighbour (NNN) Heisenberg exchange terms, namely $J_1$ and $J_2$, respectively. However, we mention that the in-plane DMI components ($D_{x,y}$) being negligibly small due to the dominant out-of-plane DMI ($D_z$) in most kagome materials~\cite{Lee1997,Zorko2008,Inami2001,Cepas2008} we drop the in-plane DMI terms here. To inspect a frustrated magnetism on our system, the noncoplanarity or the canted configuration (away from $120^\circ$ N'eel ordering) of KAFM necessitates an externally applied out-of-plane magnetic field that solely generates a nontrivial magnon topology and MTHE~\cite{Owerre2017}. In general, the interaction between quasiparticles in such a KAFM model can be intricate and hence requires involved techniques to derive an effective magnon picture. Here, we wish to investigate how the onsite (local) magnon modes couple to the optical phonon modes, resulting in a magnon-polaron state, and examine its impact on the magnon topology and the relative thermal Hall features of the entire KAFM structure.
Focusing on this primary goal, we first consider that the optical phonon modes can couple with the spin (magnon) modes through two distinct mechanisms: (i) magnon modes at a particular lattice site interact with the same onsite phonon modes~\cite{Weisse1999}, named \textit{\text{``local spin-phonon''}} coupling with a strength $g^l$, and (ii) local magnon interaction that involves phonons present at two NN lattice sites~\cite{Weisse1999,Chen2021,Go2019}, dubbed \textit{\text{``non-local spin-phonon''}} coupling with a strength $g^{nl}$. The purpose of dealing with optical (rather than the acoustic) phonon branches is to incorporate a stronger perturbation on the local atomic environment, and hence it ascertains more efficient coupling between the spin excitations and phonons~\cite{Chen2021}, which can be probed via photo-excitation experiments. Furthermore, going beyond the adiabatic (high frequency) regime enables phonons to acquire a finite energy that stabilizes the KAFM configuration even at zero temperature, although a constraint on the phonon frequency should be imposed to avoid instability of the magnon-polaron bands. In addition, as optical modes are largely non-dispersive, it is convenient to model the phonon bands generically at small momenta~\cite{Seifert2024} that are relevant for gap closing transitions of the KAFM bands. These justify the rationale for taking momentum-independent optical phonons into account. 
However, analytical methods for tackling such a coupling in a KAFM have not been proposed yet. Therefore, here we provide an analytical approach in the spirit of the spin-Peierls transformation method~\cite{Weisse1999} to yield a finite temperature magnon-polaron hybridized bound state by completely decoupling these two distinct couplings on a KAFM, and then analyze numerically its topological properties and the thermal Hall conductivity. Although hitherto proposals on the kagome ferromagnet~\cite{Sheikhi2021} and the kagome spin ice~\cite{Wu2025} have explained the effects of phononic excitations, the salient topology of the magnon-phonon hybridization in a frustrated geometry, such as a KAFM, has not been addressed till now. Therefore, the present study may hint at a richer topological scenario and bridge the gap between frustrated magnetism and magnon-polaron quasiparticle physics. 
Thus far, the following pertinent questions arise: (i) Can a topological transition be induced on the KAFM magnon-polaron bands solely by tuning the spin-phonon coupling strength? (ii) How will the magnon-polaron band topology differ in local and non-local interaction regimes, both qualitatively and quantitatively? (iii) What will the response of the topological thermal Hall conductivity be to such magnon-polaron hybridization in both these regimes?

To address these points, we organize the remainder of the paper as follows. We begin by mathematically formulating the non-coplanar KAFM magnon model, and construct the momentum ($k$) space magnon-polaron Hamiltonian, followed by the exact decoupling of the aforementioned coupling through a canonical transformation in Sec.~\ref{Formalism}. Subsequently, in Sec.~\ref {results}, we describe the numerical results in the following sequence. The bulk spectral properties accompanied by relative transitions in the Chern number are examined in Sec.~\ref{results:bulk} via varying both the local and non-local coupling strengths distinctly, while the edge state features of the topological magnon-polaron bands are analyzed in Sec.~\ref{results:edge}. Thereafter, the nature of the topological thermal Hall conductivities exhibiting the transitions is elaborately discussed in Sec.~\ref{results:THE}. Finally, we summarize and conclude our findings in Sec.~\ref{conclusion}.     

\section{Magnon-polaron model on kagome antiferromagnet}
\label{Formalism}
Let us formulate the total Hamiltonian of a tripartite kagome structure (presented in Fig.~\ref{Kagome}(a)) comprising three terms, which can be expressed as
\begin{equation}
    H = H_\text{s} + H_\text{ph} + H_\text{sp} . \label{Magnon-phonon-Hamiltonian}
\end{equation}
Here, $ H_{\text{s}} $ represents the Hamiltonian for magnetic interactions of different kinds, $ H_{\text{ph}} $ denotes the Hamiltonian for the phonon modes, and $ H_{\text{sp}} $ corresponds to that of the spin-phonon interaction considered in our system. In the following discussion, we describe each of these terms in detail and hence obtain an effective Hamiltonian to compute the topological properties and thermal Hall signatures.

\subsection{Magnon Hamiltonian}
We consider several magnetic interactions with antiferromagnetic spin configurations in $H_\text{s}$, which can be written as
\begingroup
\allowdisplaybreaks
\begin{equation}
\begin{split}
    H_\text{s} &= J_1 \sum_{{\braket{i,j}}} \left(\mb{S}_i \cdot \mb{S}_j\right) + D \sum_{{\braket{i,j}}} \nu_{ij}\hat{\mb{z}} \cdot \left(\mb{S}_i \times \mb{S}_j\right)\\
    &+ J_2 \sum_{\braket{\braket{i,j}}} \left(\mb{S}_i \cdot \mb{S}_j\right) -\mu_B B_0 \sum_i S_{zi},
\end{split}\label{Spin_Hamiltonian_kagome}
\end{equation}
where $J_1 (>0)$ and $J_2 (>0)$ are the NN and the NNN antiferromagnetic exchange interaction strengths, as mentioned earlier. Additionally, $D$ is the coefficient of an out-of-plane DMI acting between two neighbouring spins, where $\nu_{ij}=\mp 1$ depending on their lattice positions, that is, negative (positive) for an upper (lower) triangle (see Fig.~\ref{Kagome}(b))~\cite{Owerre2017}. Finally, the Zeeman term involves the external transverse magnetic field $B_0$ that eventually generates the canted orientations of spins and further aids in inducing the nontrivial topology in the system. The lattice vectors connecting two same sublattice sites are denoted by, $\delta_1 = a_0(1,\sqrt{3})$, $\delta_2 = a_0(1,-\sqrt{3})$, $\delta_3 = a_0(2,0)$.
\endgroup

Further, we consider non-collinear spin configurations with antiferromagnetic environment, where, neglecting all the spin fluctuations, the classical spin vectors at three different sublattice sites (A, B, C, as displayed in Fig.~\ref{Kagome}(a)) are given by
\begin{equation}
    \begin{split}
        \mb{S}_A &= -\frac{\sqrt{3}}{2}\cos{\eta}~\hat{\mb{x}} - \frac{1}{2} \cos{\eta}~\hat{\mb{y}} + \sin{\eta}~\hat{\mb{z}},\\
        \mb{S}_B &= \frac{\sqrt{3}}{2}\cos{\eta}~\hat{\mb{x}} - \frac{1}{2} \cos{\eta}~\hat{\mb{y}} + \sin{\eta}~\hat{\mb{z}},\\
        \mb{S}_C & = \cos{\eta}~ \hat{\mb{y}} + \sin{\mb{\eta}}~\hat{\mb{z}}.
    \end{split}\label{spin_orientation}
\end{equation}
Here, $\eta$ denotes the canting angle between the spin vectors and the lattice plane, as shown in Fig.~\ref{Kagome}(c). In the absence of an external magnetic field, the spin configuration exhibits a $120^\circ$ N\'eel (coplanar) order, with $\eta = 0$. However, in the presence of an external magnetic field ($B_0 \neq 0$), $\eta$ acquires a finite value, which can be determined by minimizing the classical energy ($\partial E_{\text{cl}}/\partial \eta = 0$) corresponding to $H_\text{s}$ (Eq.~\eqref{Spin_Hamiltonian_kagome}). This yields the relation $\sin{\eta} = B_0 / B_s$, with $B_s$ being the saturation magnetization given by, $B_s = 2\sqrt{3}[D + \sqrt{3}(J_1 + J_2)]$, corresponding to the ground-state spin configuration as reported in literature~\cite{Owerre2017,Laurell2018}.

Due to this canted spin orientation, the spin quantization axes at different sublattice sites not only differ from one another but also deviate from the global $z$-axis. Therefore, to establish a relation between the global spin operator $\mb{S}_i$ and the local spin operator $\tilde{\mb{S}}_i$, we need to introduce a sublattice-dependent rotation matrix, which can be expressed as 
\begin{equation}
    \mb{S}_i = R_z(\phi_i)R_x(\theta)\mb{\tilde{S}}_i.~~~i \in A, B, C \label{rotation_kagome}
\end{equation}
Here, $R_x(\theta)$ signifies an anticlockwise rotation about the $x$-axis by an angle $\theta$, where $\theta = 90^\circ - \eta$. Similarly, $R_z(\phi_i)$ represents an anticlockwise rotation about the $z$-axis by a sublattice-dependent angle $\phi_i$, with $\phi_i = \frac{2\pi}{3}$, $-\frac{2\pi}{3}$, and $0$ for the A, B, and C sublattices, respectively. Moreover, to produce low-energy excitations (or magnons) from $H_\textbf{s}$, we employ a spin-boson transformation, which applies only to local spin operators. Specifically, we perform the linearized Holstein-Primakoff (HP) transformation~\cite{Holstein1940}, where the spin operators $\tilde{S}_i$ are expressed in terms of bosonic creation and annihilation operators ($a_i^\dagger, a_i$) as
\begin{equation}
    \tilde{S}_i^+ =\sqrt{2S}a_i,~ \tilde{S}_i^- =\sqrt{2S}a_i^\dagger,~ \tilde{S}_i^z = S - a_i^\dagger a_i. \label{HP_transformation_Kagome}
\end{equation}
Here  $\tilde{S}_i^+= \tilde{S}_x + i \tilde{S}_y$, $\tilde{S}_i^-= \tilde{S}_x - i \tilde{S}_y$ and $a_i^\dagger (a_i)$ is the creation (annihilation) operator corresponding to the A sublattice sites. Similarly, for the B and C sublattice sites, the creation (annihilation) operators are denoted by $b_i^\dagger$ ($b_i$) and $c_i^\dagger$ ($c_i$), respectively. In our case, these operators follow the same transformation as given in Eq.~\eqref{HP_transformation_Kagome}. In the frustrated antiferromagnetic system, the spin orientations at the different sublattice sites are specified in Eq.~\eqref{rotation_kagome}. By applying the spin-boson transformation defined in Eq.~\eqref{HP_transformation_Kagome} to the spin Hamiltonian in Eq.~\eqref{Spin_Hamiltonian_kagome}, the Hamiltonian $H_\text{s}$ can be expressed in the magnon basis, and is given by
\begingroup
\allowdisplaybreaks
 \begin{eqnarray}
      H_\text{s}&=& S\sum_{r}\biggl[4(J_1+ J_2)\mathcal{E} + 4D{\mathcal{E}}_1 + \mu_0 B_0 \sin{\eta}\biggr]\nonumber\\
      &&\left(a_r^\dg a_r+b_r^\dg b_r + c_r^\dg c_r\right)+\biggl[J_1\!(P\! +\! iQ)\! -\! D (P_1 \!+\! iQ_1)\biggr]\nonumber\\
      &&(c_r^\dagger a_r + c_r^\dagger a_{r + \delta_1}
      + b_r^\dagger c_r + b_r^\dagger c_{r + \delta_2} + a_r^\dagger b_r + a_{r + \delta_3}^\dagger b_r)\nonumber\\
      &&+\biggl[J_2(P + iQ)\biggr](a_r^\dagger b_{r -\delta_2} + a_{r + \delta_1}^\dagger b_r +b_{r + \delta_1}^\dagger c_r\nonumber\\
      &&~~~~~~~~~~~~~~~~~~~~~~~~~~+ b_r^\dagger c_{r + \delta_3} + c_{r + \delta_2}^\dagger a_r + c_r^\dagger a_{r + \delta_3})\nonumber\\
      &&+\biggl[J_1 R - D R_1\biggr](c_r^\dagger a_r^\dg  + c_r^\dagger a_{r + \delta_1}^\dg + b_r^\dagger c_r^\dg  + b_r^\dagger c_{r + \delta_2}^\dg\nonumber\\  
      &&~~~~~~~~~~~~~~~~~~~~~~~~~~~~~~~~~~~~~~~~~~~~+ a_r^\dagger b_r^\dg  + a_{r + \delta_3}^\dagger b_r^\dg)\nonumber\\
      &&+J_2R(a_r^\dagger b_{r -\delta_2}^\dg  + a_{r + \delta_1}^\dg  b_r^\dg  +b_{r + \delta_1}^\dagger c_r^\dg + b_r^\dagger c_{r + \delta_3}^\dg \nonumber\\
      &&~~~~~~~~~~~~~~~~~~~~~~~~~+ c_{r + \delta_2}^\dg  a_r^\dg  + c_r^\dagger a_{r + \delta_3}^\dg) + h.c.,
      \label{Ham:magnon}
 \end{eqnarray}
\endgroup
where $r$ denotes the position of a unit cell and the sum runs over all the unit cells. The operators $a_r^\dg$ $(a_r)$, $b_r^\dg$ $(b_r)$ and $c_r^\dg$ $(c_r)$ are the creation (annihilation) operators corresponding to sublattices A, B, and C, respectively, located at position $r$, corresponding to the triangle highlighted in yellow in Fig.~\ref{Kagome}(a). Moreover, each unit cell contains six bonds, comprising one upper and one lower triangle associated with that unit cell. Further, the remaining quantities of Eq.~\eqref{Ham:magnon} are defined as
 \begingroup
 \allowdisplaybreaks
     \begin{align}
         &P = -\frac{1}{4}- \frac{1}{4}\sin^2\eta + \frac{1}{2}\cos^2\eta,~~
         Q = \frac{\sqrt{3}}{2} \sin{\eta},\nonumber\\
         &P_1 = \frac{\sqrt{3}}{4}\left(1 + \sin^2\eta\right), ~~Q_1 = \frac{1}{2}\sin{\eta},\nonumber\\
         &R=-\frac{1}{4} +\frac{1}{4}\sin^2\eta - \frac{1}{2}\cos^2\eta,~~R_1 = \frac{\sqrt{3}}{4}\left(1 - \sin^2\eta\right),\nonumber\\
         &\mathcal{E} = -\sin^2\eta + \frac{1}{2}\cos^2\eta,~~\mathcal{E}_1 = \frac{\sqrt{3}}{2}\cos^2\eta.
         \label{P_Q_value}
     \end{align}
We now add the phonon contribution to the magnon Hamiltonian and establish the hybridized scenario of the magnon-phonon coupled effective Hamiltonian as follows.
 
\subsection{Spin-phonon Hamiltonian} 
To incorporate the phonon environment, we now introduce the second term in Eq.~\eqref{Magnon-phonon-Hamiltonian}, which corresponds to the phonon energy and is given by
 \endgroup
\begin{equation}
    H_\text{ph} = \hbar \omega \sum_i \tilde{b}_i^\dagger \tilde{b}_i, \label{Phonon_energy}
\end{equation}
where $\omega$ is the frequency of the longitudinal optical (LO) phonons and $\tilde{b}_i^\dagger$ $(\tilde{b}_i)$ denotes the phonon creation (annihilation) operator of the $i$-th lattice site. It is important to note that, as the LO frequency is reasonably high, it denotes the largest energy scale for our studies. 

Now, we turn to the final term in Eq.~\eqref{Magnon-phonon-Hamiltonian}, which captures the spin-phonon coupling. As discussed earlier, spin-phonon coupling in a magnetic system can be modeled in various ways, depending on its underlying origin. Here, we consider that the phonon modes are strongly coupled to the local (onsite) magnon modes and inspect how this coupling affects the overall quasiparticle excitonic bands. 
To model such a coupling between two bosonic quasiparticles, we begin with a spin-phonon interaction Hamiltonian, where the phononic degrees of freedom couple to the spin exchange (Heisenberg) terms in two distinct ways, each arising from a different physical source. They are the local and the non-local terms modeled as
\begin{subequations}
\begin{align}
    H_\text{sp}^l &= g^l \sum_{\langle i,j \rangle} \left( \tilde{b}_i^\dagger + \tilde{b}_i \right) \mathbf{S}_i \cdot \mathbf{S}_j,\label{H_sp_1}\\
    H_\text{sp}^{nl} &= g^{nl} \sum_{\langle i,j \rangle} \left[\left( \tilde{b}_i^\dagger + \tilde{b}_i \right)- \left( \tilde{b}_j^\dagger + \tilde{b}_j \right)\right]\mathbf{S}_i \cdot \mathbf{S}_j,\label{H_sp_2}
\end{align}\label{H_sp_12}
\end{subequations}
\hspace{-3pt}where $H_\text{sp}^l$ represents the local interaction with a single harmonic degree of freedom at a particular site $i$ that directly modifies the nearest-neighbour exchange integral with $g^{l}$ being the coupling strength. This type of spin-phonon coupling has been previously studied in the context of the spin-Peierls transition. In contrast, $H_\text{sp}^{nl}$ designates a non-local coupling, in which the NN exchange integral depends on the lattice spacing between the two neighbouring spins, or in other words, the difference between the phononic displacements ($\tilde{b}+\tilde{b}^\dg$) separated by a NN lattice distance, where $g^{nl}$ is the corresponding coupling strength. As a result, this form of the spin-phonon coupling is more sensitive to the underlying lattice geometry than that of the local one ($g^l$). Due to this boson-boson interaction between the individual magnon excitonic mode and the intrinsic phononic vibrational mode, the surrounding lattice of each magnon mode gets distorted (similar to the electron-phonon interaction in a fermionic lattice), giving rise to a polarization potential. Consequently, analogous to the polaronic scenario of an electronic system, a magnon-polaron quasiparticle bound state can be formed, which captures the essential physics of the resultant hybridized state without altering any underlying symmetry of the system.    
\subsection{Magnon-phonon hybridization: canonical transformation approach}
To form such a hybridized spectrum, we resort to the magnon generation by Eq.~\eqref{HP_transformation_Kagome} for the $\mathbf{S}_i \cdot \mathbf{S}_j$ terms in Eq.~\eqref{H_sp_12} and employ a canonical transformation thereafter with the generators $\small{R^l= \frac{g^{l}}{\hbar \omega} \sum_{\langle i,j \rangle} \left( \tilde{b}_i^\dagger - \tilde{b}_i \right) \mathbf{S}_i \cdot \mathbf{S}_j},$ and $\small{R^{nl}= \frac{g^{nl}}{\hbar \omega} \sum_{\langle i,j \rangle}\left[\left(\tilde{b}_i^\dagger - \tilde{b}_i \right)- \left( \tilde{b}_j^\dagger - \tilde{b}_j \right)\right]\mathbf{S}_i \cdot \mathbf{S}_j}$ for $H_\text{sp}^l$ and $H_\text{sp}^{nl}$, respectively, to decouple the magnon and phonon modes (for the detailed derivation see Appendix~\ref{effective Spin-phonon coupling}). Subsequently, by taking a finite-temperature phonon averaging of the transformed Hamiltonian, $H$ in Eq.~\eqref{Magnon-phonon-Hamiltonian} can be reduced to $\widetilde{H}_R$ with the renormalized factors containing coupling strengths (as derived in Appendix~\ref{renormalized Hamiltonian}). This decoupling scheme is extensively used in the context of electron-phonon interaction~\cite{Islam2024,Bhattacharyya2024} (in the form of a coherent Lang-Firsov generator), and also selectively in spin models via spin-Peierls transformation~\cite{Weisse1999}. Therefore, we obtain a coherent state that is hybridized by magnon and phonons, where the corresponding renormalized Hamiltonian in the $k$-space, named \textit{magnon-polaron Hamiltonian}, can be written in Bogoliubov–de Gennes form using the hybridized quasiparticle basis $(\Psi^a_\mb{k}, \Psi^b_\mb{k}, \Psi^c_\mb{k}, \Psi^{a\dg}_{-\mb{k}}, \Psi^{b\dg}_{-\mb{k}}, \Psi^{c\dg}_{-\mb{k}})^T$ as
\begin{equation}
    \widetilde{H}_R(\mb{k})=\begin{bmatrix}
                            A(\mb{k}) & B(\mb{k})\\
                            B^*(-\mb{k}) & A^*(-\mb{k})
                        \end{bmatrix},\label{H_R_kspace}
\end{equation}
where the submatrices $A(\mb{k})$ and $B(\mb{k})$ can be expressed as
\begin{equation}
\begin{split}
    A(\mb{k})&=\begin{pmatrix}
                M & f_1(\mb{k}) & f_2(\mb{k})\\
                f_1^*(\mb{k}) & M & f_3(\mb{k})\\
                f_2^*(\mb{k}) & f_3^*(\mb{k}) & M
    \end{pmatrix},\\
    B(\mb{k})&= \begin{pmatrix}
                 0 & g_1(\mb{k}) & g_2(\mb{k})\\
                 g_1(-\mb{k}) & 0 & g_3(\mb{k})\\
                 g_2(-\mb{k}) & g_3(-\mb{k}) & 0
    \end{pmatrix},      
\end{split}
\end{equation}
and all the components in $A(\mb{k})$ and $B(\mb{k})$ are given below as (see Appendix \ref{Appendix: Magnon polaron Hamiltonian with spin-phonon coupling})
\begingroup
\allowdisplaybreaks
    \begin{align}
        M &= 4(J_1 + J_2)\,{\mathcal{E}} + 4D\,{\mathcal{E}}_1 + \mu_0 B_0 \sin{\eta} - \Delta^{l(nl)},\nonumber\\
        f_1(\mb{k}) &= [J_1(P + iQ) - D(P_1 + iQ_1)]e^{-\lambda_1^{l(nl)}} (1 + e^{ik\delta_3})\nonumber\\
        &~~~+ J_2 (P + iQ) e^{-\lambda_3^{l(nl)}}(e^{ik\delta_2} + e^{ik\delta_1}),\nonumber\\
        f_2(\mb{k}) &= [J_1(P - iQ) - D(P_1 - iQ_1)]e^{-\lambda_1^{l(nl)}} (1 + e^{ik\delta_1})\nonumber\\
        &~~~+ J_2 (P - iQ) e^{-\lambda_3^{l(nl)}}(e^{-ik\delta_2} + e^{ik\delta_3}),\nonumber\\
        f_3(\mb{k}) &= [J_1(P + iQ) - D(P_1 + iQ_1)]e^{-\lambda_1^{l(nl)}} (1 + e^{-ik\delta_2})\nonumber\\
        &~~~+ J_2 (P + iQ) e^{-\lambda_3^{l(nl)}}(e^{ik\delta_1} + e^{-ik\delta_3}),\nonumber\\
        g_1(\mb{k}) &= (J_1 R - D R_1)e^{-\lambda_2^{l(nl)}}(1 + e^{ik\delta_3})\nonumber\\
        &~~~+J_2 R e^{-\lambda_4^{l(nl)}}(e^{ik\delta_2} + e^{ik\delta_1}),\nonumber\\
        g_2(\mb{k}) &= (J_1 R - D R_1)e^{-\lambda_2^{l(nl)}}(1 + e^{ik\delta_1})\nonumber\\
        &~~~+J_2 R e^{-\lambda_4^{l(nl)}}(e^{-ik\delta_2} + e^{ik\delta_3}),\nonumber\\
        g_3(\mb{k}) &= (J_1 R - D R_1)e^{-\lambda_2^{l(nl)}}(1 + e^{-ik\delta_2})\nonumber\\
        &~~~+J_2 R e^{-\lambda_4^{l(nl)}}(e^{ik\delta_1} + e^{-ik\delta_3}),\label{component_Hamiltonian}
    \end{align}
\endgroup
where $l$ and $nl$ symbolize cases where the system exhibits local and non-local magnon-phonon interactions, respectively. Moreover, all the symbols and parameters appearing in this expression are defined in Appendix~\ref{Appendix: Magnon polaron Hamiltonian with spin-phonon coupling}. Further, $\widetilde{H}_R(\mathbf{k})$ follows generalized eigenvalue problem. Hence, to solve it, we introduce a non-Hermitian matrix $\widetilde{H}'_R(\mathbf{k})= \sigma_3 \widetilde{H}_R(\mathbf{k})$, where $\sigma_3 = \sigma_z \otimes I_3$ is a metric, with $\sigma_z$  being the $z$-component of the Pauli matrices and $I_3$, an identity matrix. Similar to the magnons, the magnon-polaron hybridized states also obey the Bose-Einstein (BE) statistics. Further, we only focus on the three positive energy spectra among the possible six of them. 

In the following discussion, we consider both types of couplings separately. Usually, the coherent decoupling method mentioned above works well in the high-frequency (anti-adiabatic) regime as the interaction potential deals with the LO phonons. However, at sufficiently large frequencies, the term $\Delta^{l(nl)}$ in Eq.~\eqref{component_Hamiltonian} becomes dominant and leads to imaginary eigenvalues in the energy bands for both the coupling mechanisms, which eventually indicates that the initial ground state is no longer an energy minimum state. This is an artefact of the strong spin-phonon coupling, where the LO phonons significantly affect the renormalized energy bands, and hence have a considerable impact on the classical ground state. Thus, depending on the coupling strength ($g^{l(nl)}$) and phonon frequency ($\omega$), the ground state may either be stable or unstable, while the latter scenario is termed as \textit{magnon-polaron instability}.


\begingroup
\allowdisplaybreaks

\section{Numerical analyses of magnon-polaron topology on KAFM}
\label{results}
As mentioned earlier, due to the presence of extensive degeneracy in the energy bands, a frustrated kagome antiferromagnet is a strong candidate as a QSL. However, in the presence of $J_2>0$ and a nonzero out-of-plane DMI ($D\neq 0$), the degeneracy can be lifted, stabilizing the $\bm {Q}=0$, $120^\circ$ coplanar spin ordering (as given in Eq.~\eqref{spin_orientation}), which manifests an LRO phase. Additionally, applying an out-of-plane magnetic field causes the spins to cant toward the field, resulting in the emergence of topological magnon bands and non-zero thermal Hall conductivity~\cite{Owerre2017}.
Further, in such a system, strong spin–phonon interaction involving optical phonons, whose energy is comparable to the exchange interaction strength $J_1$ can give rise to magnon-polaron hybridized quasiparticle states that retain the topological characteristics of the system. Moreover, we shall show that, without disturbing the classical ground state described by Eq.~\eqref{spin_orientation}, the spin–phonon coupling strength $g^{l(nl)}$ can drive topological phase transitions associated with these hybridized states. Additionally, its interplay with the out-of-plane DMI produces interesting results. 

As stated earlier, the presence of strong spin–phonon coupling may lead to magnon-polaron instability at larger frequencies, that is $\omega >\omega_c$ ($\omega_c$ being the critical phonon frequency). Hence, to avoid this issue, we need to analyze the magnitude of the critical frequency. Since, the band energy is being modified due to the effects of spin-phonon coupling strength and the phonon frequency (see Appendix~\ref{Appendix: Magnon polaron Hamiltonian with spin-phonon coupling}), we numerically estimate the corresponding values of critical phonon frequency $\omega_c$ (above which the instability occurs) as a function of the coupling strength $g^{l(nl)}$, while keeping all the other parameters fixed at $B_0 = 0.4B_s$, $J_2=0.03J_1$, $D=0.045J_1$, $T=0$, as shown in Fig.~\ref{omega_c_with_g}.
These parameter values are employed in most of the subsequent analyses, and it should be noted that $\omega_c$ is insensitive to minor changes in the choices of $J_2$ and $D$. Additionally, we observe that the non-local coupling mechanism is more sensitive to the phonon frequencies, as considerably lower values of $\omega_c$ can induce instability, as compared to the local one. Hence, we fix phonon energy as $\hbar \omega = J_1S$ throughout this study for consistency, as for both the coupling mechanisms $\hbar \omega_c$ is greater than $J_1S$. Moreover, this particular value of $\hbar \omega$ (in fact any value in its vicinity) simultaneously aids in avoiding the adiabatic regime (maintaining $\hbar\omega\not\ll J_1S$) and eliminates the possibility of magnon-polaron instability.
\begin{figure}
    \centering
    \includegraphics[width=1\linewidth]{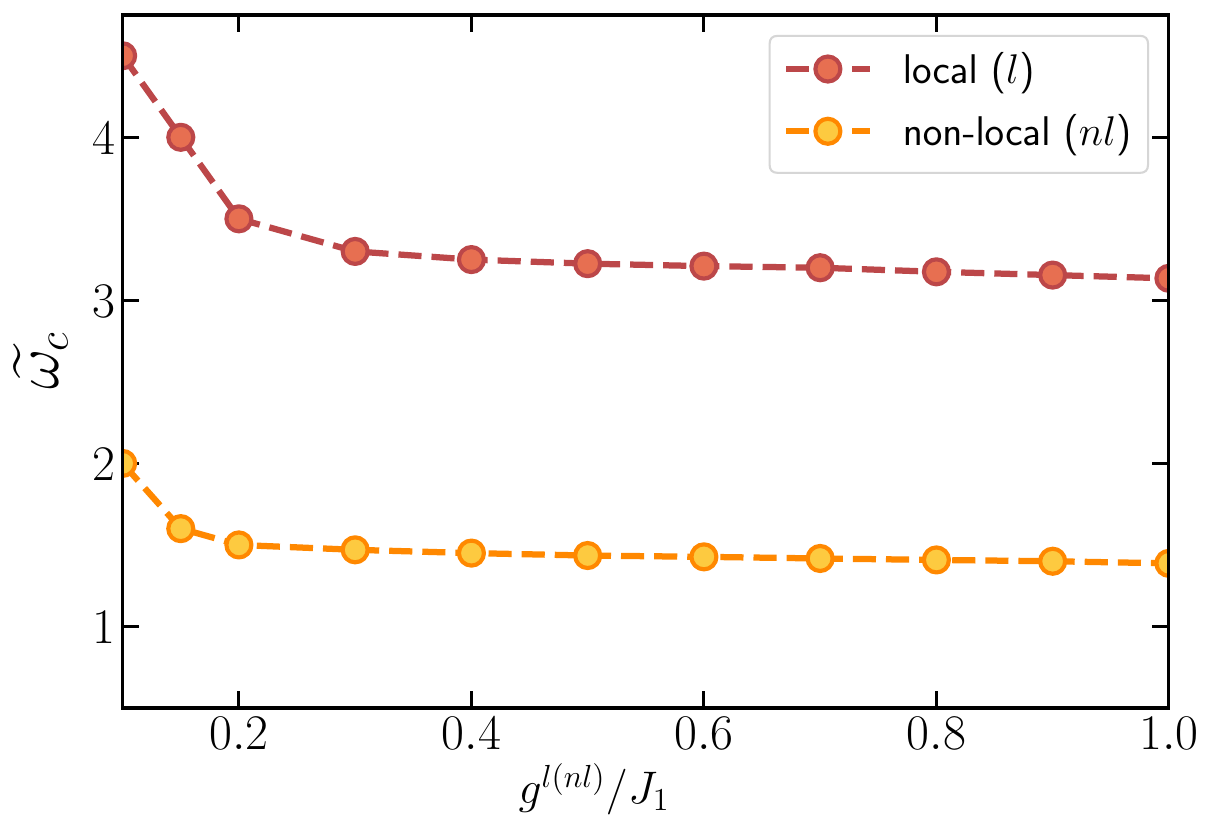}
    \caption{The behavior of the critical frequency ($\omega_c$) as a function of both local and non-local coupling strengths $g^{l(nl)}$ is shown, with $\widetilde{\omega}_c=\hbar \omega_c/J_1 S$. The other parameters are fixed at $B_0 = 0.4B_s$, $J_2 = 0.03J_1$, $D=0.045J_1$, $T=0$.}
    \label{omega_c_with_g}
\end{figure}

\subsection{Bulk spectra and topological transition by spin-phonon coupling}
\label{results:bulk}
\begin{figure*}[t]
    \centering
    \includegraphics[width=1\linewidth]{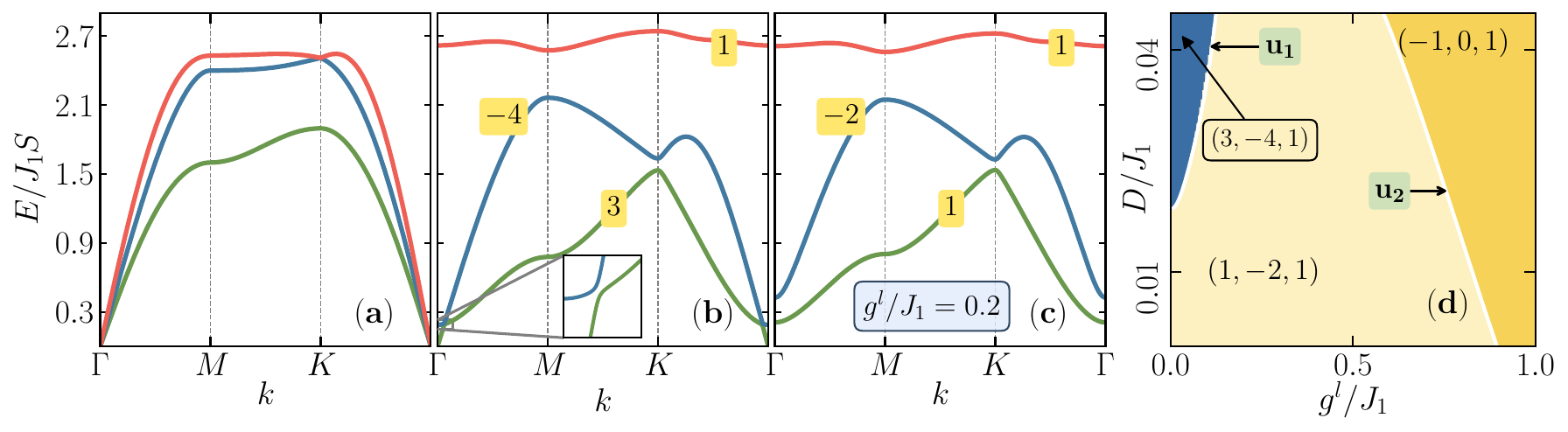}
    \caption{The bulk band structures are shown at the symmetry points inside the Brillouin zone. In the absence of any spin-phonon coupling, the bulk band structures for (a) $B_0 = D =0$, $J_2 = 0.2J_1$ and (b) $B_0 = 0.4B_s$, $J_2 = D =0.03J_1$ are plotted. (c) In the presence of local spin-phonon coupling with $\hbar \omega = J_1S$, $g^l = 0.2J_1$, $B_0 = 0.4B_s$, $J_2 = D = 0.03J_1$ the bulk band structure is shown at zero temperature. In (b) and (c), the numbers highlighted in yellow correspond to the Chern numbers. (d) A combined phase diagram based on the Chern number sets of corresponding bands is depicted in the $g^l$-$D$ plane, where the topological transitions are demonstrated explicitly.}
    \label{phase_plot_local}
\end{figure*}
While our primary motivation is to explore the topological behavior of magnon-polaron states and its effect on the thermal Hall conductivity, for the sake of a comprehensive understanding, it is first important to discuss the origin of the LRO phase in such a frustrated kagome antiferromagnet, as well as the key elements that enable the system to exhibit Dirac semimetal behavior.
In the absence of any long-range interactions and DMI, the energy band structure obtained from Eq.~\eqref{Spin_Hamiltonian_kagome} for the $120^\circ$ coplanar spin configuration exhibits zero-energy modes throughout the Brillouin zone (BZ), indicating a gapless spectrum between the ground and the excited states. However, upon introducing NNN interaction with $J_2>0$, the bands become dispersive and acquire finite energy along the high-symmetry points inside the BZ, such as, $\mb{\Gamma}=(0,0)$, $\mb{M} = (\frac{\pi}{2}, \frac{\pi}{2\sqrt{3}})$ and $\mb{K}= (\frac{2\pi}{3}, 0)$ as seen in Fig.~\ref{phase_plot_local}(a). Since the system preserves valley symmetry~\cite{Rehman2022}, the energies at the two Dirac points $\mb{K}$ and $\mb{K'}$ remain degenerate. Also, in Fig.~\ref{phase_plot_local}(a), all the three energy bands exhibit Goldstone modes at the $\mb{\Gamma}$ point due to the presence of SO(3) rotational symmetry, where it is reduced to SO(2) symmetry due to the presence of an out-of-plane DMI, resulting in only one Goldstone mode occuring at the $\mb{\Gamma}$ point~\cite{Owerre2017}. Furthermore, the middle and the upper bands cross linearly at the Dirac point $\mb{K}$, which makes the system a Dirac semimetal.

Since the TRS operator $\tau$ flips the spin direction, any magnetically ordered system violates TRS and instead follows an effective TRS. Here, for this $120^\circ$ N\'eel order, the effective TRS is the combination of TRS and the mirror symmetry with respect to the kagome plane. While out-of-plane DMI could not break this symmetry, an external magnetic field ($B_0$) causes the spins to be canted along the field direction, which breaks effective TRS. Further, from Eqs.~\eqref{P_Q_value} and~\eqref{component_Hamiltonian}, it can be shown that in the absence of an external magnetic field, as long as $\eta = 0$, the $\mb{k}$-independent imaginary part of the hopping strength remains zero. However, for a nonzero value of $\eta$, the Hamiltonian breaks time-reversal symmetry, that is, $\widetilde{H}_R(-\mb{k}) \neq \widetilde{H}_R^*(\mb{k})$. Consequently, in the presence of both $D$ and $B_0$, we observe gapped magnons everywhere inside the BZ with one Goldstone mode occurring at $\mb{\Gamma}$, as shown in Fig.~\ref{phase_plot_local}(b). 

\begingroup
Further, to determine whether these band gaps are topological or trivial in such a TRS-broken system, one can compute the Chern numbers of the corresponding bands. The Chern number ($C_n$) associated with $n^{\text{th}}$ band is given by~\cite{thouless1998topological}
\begin{equation}
    C_n = \frac{1}{2\pi}\int_{BZ} \mb{\Omega}_n^z(\mb{k})\cdot d\mb{k}, \label{chern}
\end{equation}
where $n$ is the band index and $\Omega_n^z(\mb{k})$ is the Berry curvature corresponding to the $n^{\text{th}}$ band. Moreover, in the presence of broken TRS, Berry curvature becomes nonzero and is given by~\cite{Debnath2025,Bhattacharyya2024}
\begin{equation}
    \mb{\Omega}_n^z(\mb{k})= -2 \text{Im} \braket{\partial_{k_x}\psi_n(\mb{k})|\partial_{k_y}\psi_n(\mb{k})}, \label{Berry_curvature}
\end{equation}
where $\psi_n(\mb{k})$ is the eigenfunction corresponding to the $n^{\text{th}}$ band. In general, the Berry curvature attains larger values near the high symmetry points ($\mb{\Gamma}$, $\mb{M}$, $\mb{K}$, $\mb{K'}$), where the dispersive bands exhibit a large curvature. Therefore, upon calculating the Chern number from Eq.~\eqref{chern}, we find that the Chern numbers for the lower, middle, and upper bands are $3$, $-4$, and $1$, respectively, as shown in Fig.~\ref{phase_plot_local}(b). 
\endgroup

\begin{figure}[t]
    \centering
    \includegraphics[width=1\linewidth]{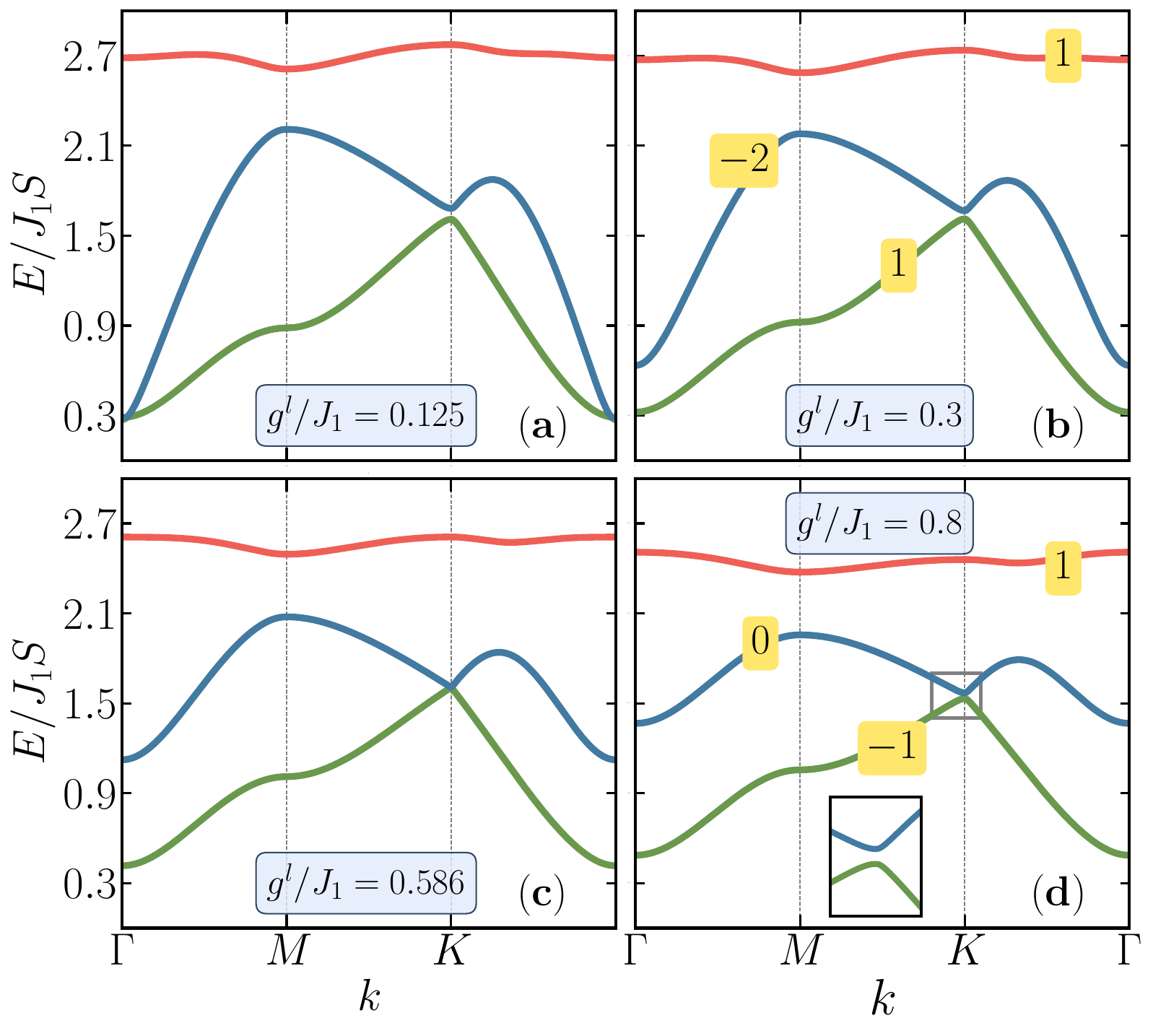}
    \caption{The bulk band structures are displayed along the high symmetry path, corresponding to the local spin-phonon coupling for $B=0.4 B_s$, $J_2 = 0.03J_1$, and $D = 0.045J_1$ with (a) $g^l = 0.125J_1$, (b) $g^l = 0.3J_1$, (c) $g^l = 0.586J_1$, (d) $g^l = 0.8J_1$. The numbers highlighted in yellow denote the Chern numbers corresponding to the bands shown in (b) and (c).}
    \label{gap_closing}
\end{figure}
\begin{figure*}[t]
    \centering
    \includegraphics[width=1\linewidth]{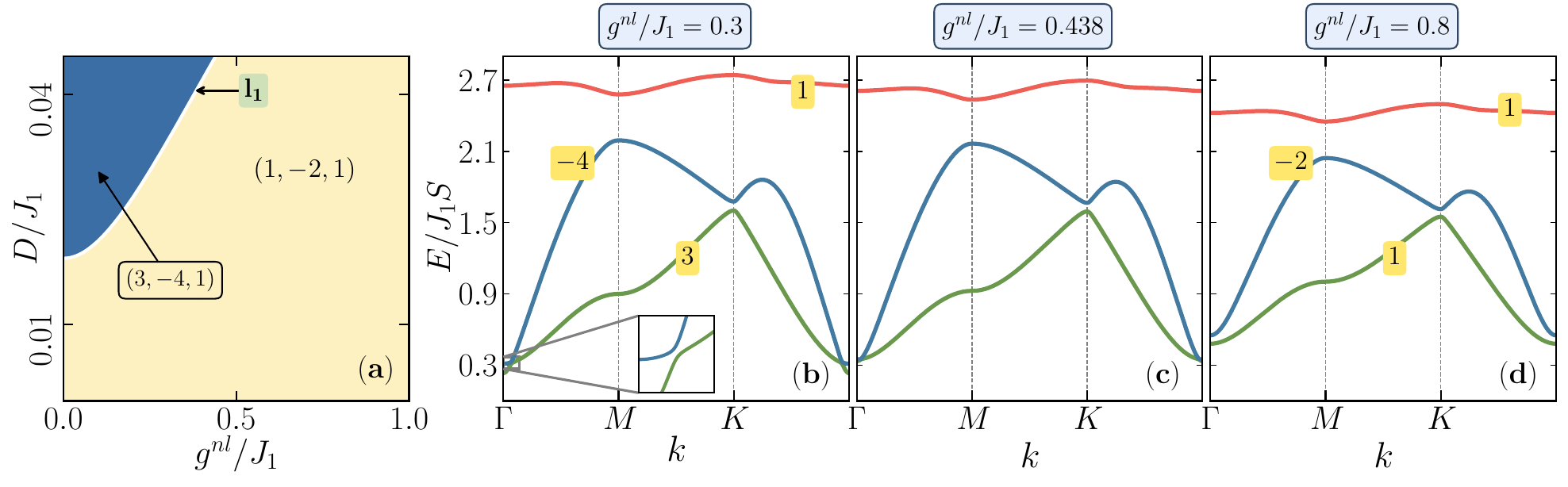}
    \caption{(a) A combined phase diagram based on the Chern number sets of the corresponding bands is illustrated as a function of the non-local coupling $g^{nl}$ and the out-of-plane DMI $D$. The bulk band structures are shown along the high symmetry points with $B=0.4 B_s$, $J_2 = 0.03J_1$, $D = 0.045J_1$, $\hbar \omega = J_1S$ for (b) $g^{nl} = 0.3J_1$, (c) $g^{nl} = 0.438J_1$, and (d) $g^{nl} = 0.8J_1$. The Chern numbers corresponding to the bands are denoted in yellow. These figures ascertain the occurrence of topological transitions upon solely tuning $g^{nl}$.}
    \label{Phase_plot_nonlocal}
\end{figure*}
Furthermore, keeping all parameter values the same as in Fig.~\ref{phase_plot_local}(b), the inclusion of the local spin-phonon interaction leads to the conversion of pure magnon bands into hybridized magnon-polaron states. Moreover, for a coupling strength of $g^l = 0.2J_1$, we observe that the Goldstone mode near the $\mb{\Gamma}$ point vanishes due to a positive energy shift of the lower and middle magnon-polaron bands near the $\mb{\Gamma}$ point in Fig.~\ref{phase_plot_local}(c),  the reason of which is explained below.

Here, from Eq.~\eqref{component_Hamiltonian}, we observe that in the presence of local spin-phonon interaction ($H_\text{sp}^l$), the effective hopping strengths are reduced by a factor of $e^{-\lambda_q^l}$, where $q\in 1,2,3,4$ (see Eq.~\eqref{H_R_eff}), which helps to stabilize the low-energy magnon-polaron states. On the other hand, the onsite energy is modified by a negative energy shift $\Delta^l$ proportional to $(\alpha^l)^2\omega$, which tends to destabilize the magnon-polaron states. Since our system already hosts a Goldstone mode near the $\mb{\Gamma}$ point, if the effect of $\Delta^l$ outweighs that of the $e^{-\lambda_q^l}$ term at large $\omega$, the system may slip into a regime with magnon-polaron instability, leading to imaginary eigenvalues within the BZ. Therefore, we keep $\hbar\omega = J_1S$, where the exponential factor dominates and hence avoids the magnon-polaron instability.

In addition, to ensure that the system undergoes a topological phase transition, we calculate the Chern numbers corresponding to these bands and observe a transition from $C = 3 \rightarrow C = 1$ for the lower band and from $C = -4 \rightarrow C = -2$ for the middle band, in the presence of $g^{l}$ (at $g^{l}=0.2J_1$), as shown in Fig.~\ref{phase_plot_local}(c). Hence, we observe that the presence of local spin-phonon coupling can significantly influence the topological state of the system. Furthermore, to examine the interplay between DMI and the spin-phonon coupling strength, we construct a topological phase diagram, shown in Fig.~\ref{phase_plot_local}(d). Here, distinct topological regions are denoted by the Chern number set $C = (C_1, C_2, C_3)$, where the indices $1$, $2$, and $3$ correspond to the lower, middle, and upper bands, respectively. Furthermore, in this phase diagram, we fix the other parameters at values given by $B = 0.4B_s$, $J_2 = 0.03J_1$, and set the temperature to zero ($T = 0$). Then, by assuring $D$ to be fixed and varying $g^l$ from $1$ to $J_1$, we observe multiple topological phase transitions along the $\text{u}_1$ and $\text{u}_2$ transition lines corresponding to a given classical ground state. Moreover, each of these transitions must be associated with a gap closing transition, as can be seen from spectral features.

To visualize this, we present the bulk band structure near the transition points in Fig.~\ref{gap_closing}. Despite finite values of $B_0$ and $D$, we observe a gap closing scenario near the $\mb{\Gamma}$ point between the lower and the middle bands along the transition line $\text{u}_1$ (see Fig.~\ref{gap_closing}(a)). The bands reopen beyond the $\text{u}_1$ contour, as shown in Fig.~\ref{gap_closing}(b), where the Chern number set is given by $C = (1, -2, 1)$. Furthermore, with increasing coupling strength $g^l$, we observe another gap closing event at larger values of $g^l$ near the $\mb{K}$ point in Fig.~\ref{gap_closing}(c), which becomes gapped again beyond the $\text{u}_2$ line with the Chern number set $C = (-1, 0, 1)$, as shown in Fig.~\ref{gap_closing}(d). While we keep the phases corresponding to low values of $J_2$ and $D$ (to maintain the $\bm{Q}=0$ LRO phase) in the main text, the results for relatively larger values of $J_2$ have been shown in Appendix~\ref{Appendix_Phase_diagrams}.
\endgroup

Thus far we have investigated multiple phase transitions in the presence of local spin-phonon interaction ($H_\text{sp}^l$), where the involvement of phonon degrees of freedom differs from that in the non-local spin-phonon interaction ($H_\text{sp}^{nl}$), as given in Eq.~\eqref{H_sp_12}. Although similar modifications appear in Eq.~\eqref{component_Hamiltonian} in the presence of $H_\text{sp}^{nl}$, the magnitudes of the effective parameters, such as $\lambda_q^{nl}$ and $\Delta^{nl}$, differ from those in the local case. Consequently, at $\hbar \omega = J_1 S$, the interplay between these factors yields distinct outcomes.

To provide further insight into these mechanisms, we also construct the topological phase diagram as a function of $D$ and the non-local spin-phonon coupling strength $g^{nl}$. Here, in Fig.~\ref{Phase_plot_nonlocal}(a), we observe two different topological regions with distinct Chern number sets $C = (3, -4, 1)$ and $C = (1, -2, 1)$. Further, the transition between these regions, marked by the line $\text{l}_1$, must involve a gap closing event. Moreover, prior to crossing $\text{l}_1$, a narrow gap separates the lower and middle bands near the $\mb{\Gamma}$ point, which becomes gapless along that transition line, as seen from Figs.~\ref{Phase_plot_nonlocal}(b) and \ref{Phase_plot_nonlocal}(c). However, beyond the transition line $\text{l}_1$, a gap reopens with a distinct Chern number set and the system retains its topological characteristics as shown in Fig.~\ref{Phase_plot_nonlocal}(d).

Although in both cases we observe that with increasing $g^{l(nl)}$, the band energy corresponding to the lower band increases near the $\mb{\Gamma}$ point. These characteristics depend on the competition between the factors of the renormalized magnon-polaron bands, namely $\Delta^{l(nl)}$ and $e^{-\lambda_q^{l(nl)}}$ (Appendix~\ref{renormalized Hamiltonian}) at a fixed value of the phonon frequency, $\omega$. Therefore, if we increase $\omega$, it may cause the lower band energy to decrease near the $\mb{\Gamma}$ point with increasing $g^{l(nl)}$, which may eventually acquire imaginary eigenvalues and induce a magnon-polaron instability near the $\mb{\Gamma}$ point above a certain limit of coupling strength. 
Notably, the key difference in the outcome for the two distinct couplings is observed from the phase plots, which indicates that tuning of the local coupling strength ($g^{l}$) yields an additional phase (phase diagram comprises of three distinct phases) compared to the non-local case ($g^{nl}$), which renders two distinct phases. The plausible reason behind this disparity stems from the difference in the polaronic weights ($\lambda^{l(nl)}_q$) of the magnon-polaron bands corresponding to these two cases (see Appendix~\ref{renormalized Hamiltonian}). While the renormalization due to the local coupling gives rise to more complex and dominant effects, leading to strong band-narrowing features, the non-local coupling involving NN phonon modes suffers screening and thus produces weaker effects.

\subsection{Chiral magnon-polaron edge modes scenario}
\label{results:edge}
So far, we have explored the topological characteristics and phase transitions induced solely by spin-phonon interactions through the analysis of bulk band structures and their associated Chern numbers. However, it is equally important to verify these topological features by examining the edge modes in order to establish the bulk-boundary correspondence. While the magnonic edge modes in kagome antiferromagnets under an external magnetic field, without spin-phonon interactions, have been investigated in literature~\cite{Owerre2017}, the present work focuses on hybridized magnon-polaron states rather than the bare magnons. Subsequently, to examine the bulk-boundary correspondence, we consider a two-dimensional semi-infinite ribbon geometry, finite along one direction (for our case, it is the $y$-direction) and infinite along the other ($x$-direction). Further, to analyze the number of crossings between the edge modes as well as their direction of propagation, we compute the winding number ($\mathcal{W}_n$), defined as the sum of the Chern numbers up to the $n^{\text{th}}$ band and is given by~\cite{Mook2014,Debnath2025}
\begin{figure}[t]
    \centering
    \includegraphics[width=1\linewidth]{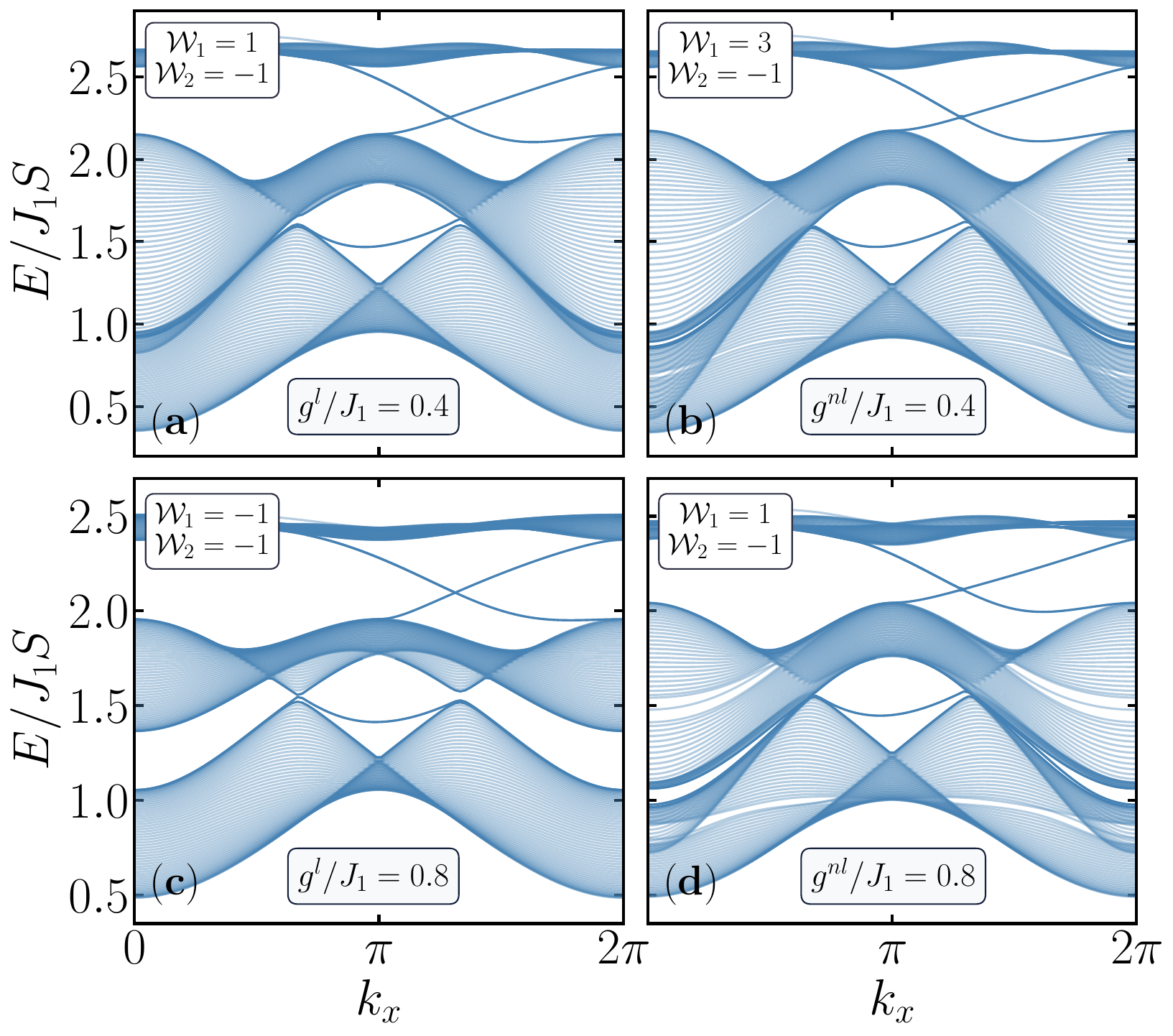}
    \caption{The edge states of the magnon-polaron bands corresponding to both types of spin-phonon interaction are shown. In the presence of local spin-phonon coupling, the edge modes are plotted for (a) $g^{l} = 0.4J_1$, and (c) $g^{l} = 0.8J_1$. Similarly, for the non-local spin-phonon coupling, the edge modes are displayed for (b) $g^{nl} = 0.4J_1$ and (d) $g^{nl} = 0.8J_1$. In all the cases, the other parameters are fixed at $B_0 = 0.4 B_s$, $J_2 = 0.03 J_1$, and $D = 0.045 J_1$ at zero temperature. These figures establish the bulk-boundary correspondence for our study.}
    \label{Edge}
\end{figure}
\begin{equation}
    \mathcal{W}_n = \sum_{i=0}^n C_i,\label{winding}
\end{equation}
where $i$ is the band index and $n$ is the gap index. The values of $\mathcal{W}_n$ give the number of crossings at the $n^{\text{th}}$ band gap, and its sign signifies the direction of propagation of the edge current or the overall chirality. A positive winding number indicates positive chirality, where the edge current propagates from left to right along the top edge and from right to left along the bottom edge. Similarly, a negative winding number has analogous implications.
In our system, we consider a ribbon consisting of 60 lattice sites along the $y$-direction, while keeping it infinite along the $x$-direction, making only  $k_x$ a good quantum number. 

Here, to study the fate of the edge states in the topologically gapped regimes of Figs.~\ref{gap_closing} and \ref{Phase_plot_nonlocal}, we compute $\mathcal{W}
_n$ for both $H_\text{sp}^l$ and $H_\text{sp}^{nl}$, using a fixed set of parameters: $B = 0.4B_s$, $D = 0.045J_1$, $J_2 = 0.03J_1$, and $T = 0$, to enable a direct comparison between the two cases. Now, in presence of $H_\text{sp}^l$, with $g^l = 0.4J_1$, the Chern number set associated with the bands is $C = (1, -2, 1)$. Therefore, according to Eq.~\eqref{winding}, we find $\mathcal{W}_1 = 1$, indicating one crossing of the edge modes in the first spectral gap, with positive chirality, and $\mathcal{W}_2 = -1$, indicating another such crossing in the second spectral gap, with negative chirality. Furthermore, in Fig.~\ref{Edge}(a), due to an overlap between the lower and the middle bands, there is no such crossing corresponding to $\mathcal{W}_1$, while another edge crossing corresponding to $\mathcal{W}_2$ is clearly visible at $g^l=0.4J_1$. Further, by increasing $g^l$ to $0.8J_1$, the system undergoes a topological phase transition, with the Chern number set changing from $C = (1, -2, 1)$ to $C = (-1, 0, 1)$. Hence, applying Eq.~\eqref{winding}, we find $\mathcal{W}_1 = -1$ and $\mathcal{W}_2 = -1$, indicating a single edge crossing in each of the spectral gaps, where the corresponding edge currents traverse with negative chirality, as shown in Fig.~\ref{Edge}(c). 

As in the case of local spin-phonon interaction, the winding numbers $\mathcal{W}_n$ can also be evaluated in the presence of non-local spin-phonon coupling as well. For $g^{nl} = 0.4J_1$, the system exhibits a topological phase characterized by the Chern number set $C = (3, -4, 1)$. This yields winding numbers $\mathcal{W}_1 = 3$ and $\mathcal{W}_2 = -1$, corresponding to the lower and the upper spectral gaps, respectively. As shown in Fig.~\ref{Edge}(b), the lower band gap is not clearly defined due to an overlap, but a single edge crossing is visible in the upper band gap with negative chirality. Upon increasing $g^{nl}$ to $0.8J_1$, the Chern numbers now get modified to $C = (1, -2, 1)$, giving $\mathcal{W}_1 = 1$ and $\mathcal{W}_2 = -1$. In this case, one edge crossing appears in the upper band gap with negative chirality, while the lower and middle bands demonstrate a gapless scenario as seen in Fig.~\ref{Edge}(d). 

 Although the number of edge crossings remains the same in all cases, except for $g^{nl} = 0.4J_1$, the direction of propagation, determined by the sign of the winding number $\mathcal{W}_n$, varies distinctly across different cases considered by us. Furthermore, even with the same spin-phonon coupling strength, the local and non-local scenarios induce different topological phases. It is also noteworthy that in both Figs.~\ref{Edge}(a) and \ref{Edge}(b), the lower and the middle bands exhibit overlap along the $k_x$, consistent with the bulk band structure, where these bands lie in close proximity of the $\mb{\Gamma}$ and the $\mb{K}$ points.

\subsection{Magnon-polaron thermal Hall conductivity}\label{results:THE}
We have seen so far that in the presence of an external magnetic field, a frustrated kagome antiferromagnet exhibits various topological phase transitions driven by strong spin-phonon coupling involving optical phonons. Although we have characterized the topological states using Chern numbers and the bulk-boundary correspondences have been robustly tested, Hall conductivity, being one of the measurable quantities in experiments, remains to be explored. Unlike fermionic systems, where the occupied states below the Fermi level contribute to the quantized Hall conductivity, in bosonic systems, all the states with positive energy can contribute to the thermal Hall conductivity. Nevertheless, at low temperatures, the dominant contribution arises from the highly occupied low-energy states. The thermal Hall conductivity (scaled by $\frac{\hbar}{k_B}$) is given by~\cite{dofhall,MatsumotoII2011}
\begin{equation}
    \kappa_{xy} = \frac{k_B T}{4 \pi^2} \sum_n \int_{BZ} c_2(\rho_{n,\mathbf{k}})\Omega^{z}_n(\mathbf{k}) \cdot d\mb{k}, \label{kappa}
\end{equation}
where $k_B$ is the Boltzmann constant, $T$ is the temperature, and $\rho_{n, \mathbf{k}} = 1/[\exp{(\varepsilon_n(\mathbf{k}) / k_B T)} - 1]$ denotes the BE distribution function, with $\varepsilon_n(\mathbf{k})$ representing the eigenvalue associated with the $n^{\text{th}}$ band. Moreover, $c_2(\rho_{n, \mathbf{k}})$ serves as the weighting function, defined as
\begin{eqnarray}
    c_2(\rho_{n,\mathbf{k}})&=&(1 + \rho_{n,\mathbf{k}})\left(\log{\frac{1 + \rho_{n,\mathbf{k}}}{\rho_{n,\mathbf{k}}}}\right)^2
    - (\log{\rho_{n,\mathbf{k}}})^2 \nonumber\\
    &&- 2\text{Li}_2 (- \rho_{n,\mathbf{k}}), 
\label{c2}
\end{eqnarray}
where $\text{Li}_2(-\rho_{n,\mathbf{k}})$ denotes the polylogarithmic function. Further, the function $c_2(\rho_{n,\mathbf{k}})$ attains larger values for lower-energy bands at low temperatures~\cite{Debnath2025}. As a result, the low-energy spectrum dominates the behavior of the thermal Hall conductivity, while contributions from the higher-energy states become insignificant.

As mentioned earlier, the LRO phase can be stabilized at small values of $J_2$ and $D$ relative to $J_1$. Moreover, while analyzing the band structures and their topological properties, the external magnetic field ($B_0$) has been fixed at $0.4 B_s$ (equivalent to $2.53 J_1$), corresponding to a specific classical ground state. However, varying the external magnetic field, while keeping all other parameters fixed at $J_2 = 0.03 J_1$, $D = 0.045 J_1$, $\widetilde{T}=0.4$ ($\widetilde{T} = k_B T/J_1 S$), and $\hbar\omega = J_1S$~\cite{3rdwork_note}, leads to a change in the classical ground state and may give rise to distinct topological phases.  

 \begin{figure*}[t]
    \centering
    \includegraphics[width=1\linewidth]{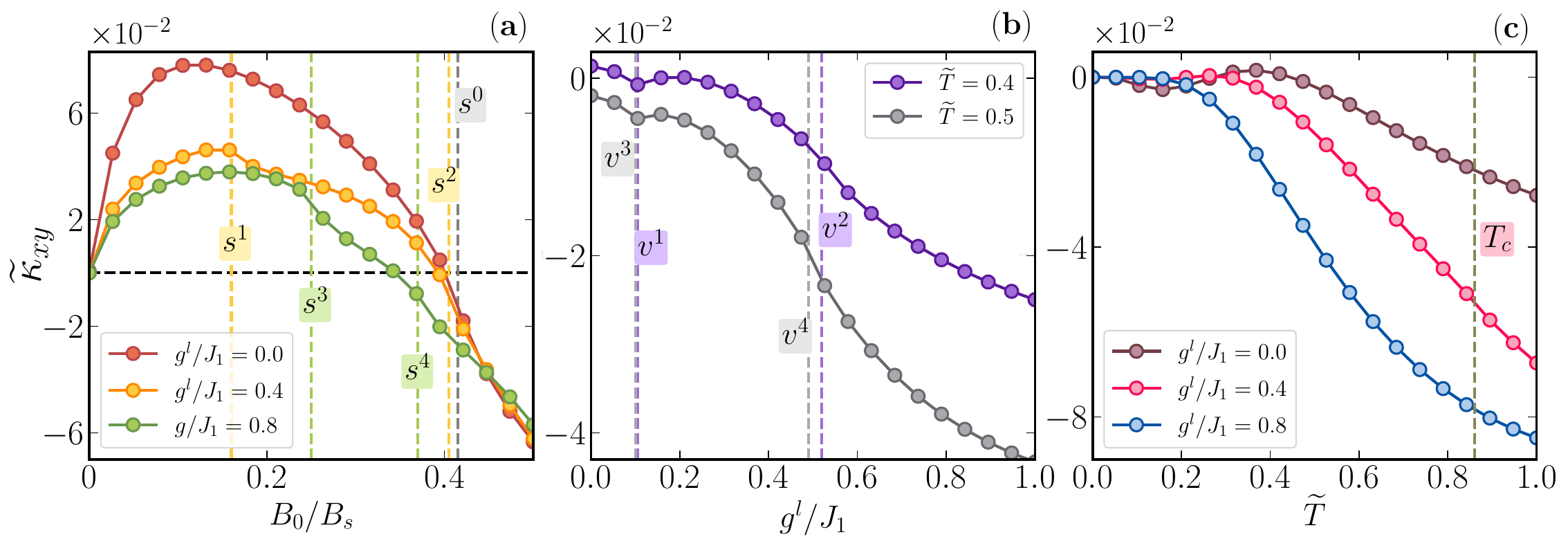}
    \caption{The effects of local spin-phonon interaction on the behavior of the thermal Hall conductivity ($\kappa_{xy}$) are presented. (a) $\kappa_{xy}$ is plotted as a function of external magnetic field for different values of the coupling strength $g^l$ at $\widetilde{T}= 0.4$. The transition lines $s^0$ correspond to $g^l=0$, while $s^1,~s^2$ denote $g^l/J_1 = 0.4$, and $s^3,~s^4$ refer to $g^l/J_1 = 0.8$. (b) The variations of $\kappa_{xy}$ are shown as a function of $g^l$ at two different temperatures, keeping $B_0 = 0.4 B_s$, where $v^1,~v^2$ denote $\widetilde{T}=0.4$ and $v^3,~v^4$ correspond to $\widetilde{T}=0.8$. (c) The same is also demonstrated as a function of temperature for distinct values of $g^l$. Here, $\widetilde{\kappa}_{xy} = \kappa_{xy}/J_1S$ and $\widetilde{T} = k_B T/ J_1 S$. Throughout this analysis, the other parameters are fixed at $J_2 = 0.03J_1$, $D = 0.045J_1$ and $\hbar \omega = J_1 S$.}
    \label{Hall_local}
\end{figure*}
In Fig.~\ref{Hall_local}(a), we observe that at low values of $B_0$, the Hall conductivity is positive, corresponding to the topological state $C = (3, -4, 1)$, and increases with increasing $B_0$. It reaches a peak near $B_0 \approx 0.1 B_s$, where the contribution from the lower band becomes significant. Beyond that, the Hall conductivity begins to decrease in spite of being in the same topological state, indicating a reduction in the spectral gap near the low-energy $\mb{\Gamma}$ point between the lower and the middle bands. Furthermore, we observe a topological phase transition near the $s^0$ transition line (at $B_0 = 0.42 B_s$), where the set of Chern numbers changes from $C = (3, -4, 1)$ to $C = (1, -2, 1)$. This transition is also associated with a reversal in the direction of the Hall current. The thermal Hall conductivity becomes zero at a critical value of $B_0$, where the contribution from the Berry curvature (along with that coming from the $c_2(\rho_{n,\mathbf{k}})$ function) vanishes. 

Now, considering local spin-phonon coupling with $g^l = 0.4J_1$, we observe two topological phase transitions as a function of $B_0$, each associated with distinct classical ground states. Moreover, the Hall conductivity also shows discontinuity near that transition line. Prior to the $s^1$ ($B_0 \approx 0.16 B_s$) transition line, where the Hall conductivity increases as a function of $B_0$, the topological state corresponds to $C = (1, 0, -1)$, which changes to $C = (1, -2, 1)$ beyond the $s^1$ transition line. Furthermore, a second phase transition occurs near the $s^2$ line (at $B_0 \approx 0.41B_s$), where the topological state changes from $C = (1, -2, 1)$ to $C = (-1, 0, 1)$. Moreover, for a different coupling strength, $g^l = 0.8J_1$, we observe a similar sequence of topological phase transitions involving the same sets of Chern numbers, but occurring at different values of the external magnetic field. Specifically, the first transition occurs near the $s^3$ transition line at $B_0 = 0.25 B_s$, followed by a second transition near the $s^4$ line at $B_0 = 0.37 B_s$. Nevertheless, it is important to mention that, in the presence of spin-phonon coupling, the band topology is no longer independent of temperature. Hence, the same transition points will not feature at different temperatures. We shall discuss this in more detail later.

However, based on the band properties and the associated Chern numbers, we have already demonstrated that spin-phonon coupling can induce topological phase transitions in the system. Therefore, in Fig.~\ref{Hall_local}(b), we present the behavior of the Hall conductivity as a function of the coupling strength ($g^l$), which reflects the signatures of topological phase transitions in its characteristics. Here, we examine the Hall conductivity as a function of $g^l$, varied from $0$ to $J_1$, at two different temperatures, while keeping the external magnetic field fixed at $B_0 = 0.4 B_s$. In Fig.~\ref{phase_plot_local}(d), we show that at $\widetilde{T}=0$, for the same values of $D$ and $J_2$ as mentioned earlier, two phase transitions occur as a function of $g^l$ within this range. Similarly in Fig.~\ref{Hall_local}(b), at a finite temperature, specifically, at $\widetilde{T} = 0.4$, we observe a phase transition from $C = (3, -4, 1)$ to $C = (1, -2, 1)$ across the $v^1$ line at $g^l = 0.105J_1$, whereas the other transition from $C = (1, -2, 1)$ to $C = (-1, 0, 1)$ occurs across the $v^2$ line at $g^l = 0.52J_1$. Owing to the temperature dependency on the energy bands, at $\widetilde{T} = 0.5$, the transitions occur at different values of $g^l$. As $g^l$ is varied from $0$ to $J_1$, the first transition takes place at the $v^3$ line for $g^l = 0.1J_1$, while the second transition occurs across the $v^4$ line for $g^l = 0.49J_1$. Additionally, with the first transition occurring at a low value of $g^l$, the gap closing takes place at the $\mathbf{\Gamma}$ point (see Fig.~\ref{gap_closing}(a)), where the energies corresponding to the lower and the middle bands are very small. Moreover, due to the presence of the $c_2(\rho_{n,\mathbf{k}})$ function, the primary contribution in the thermal Hall conductivity comes from that region of the BZ (vicinity of the $\mb{\Gamma}$ point), resulting in a noticeable kink near $v^{1(3)}$ transition line. Further, at $v^{2(4)}$, the gap closing transition occurs at the $\mathbf{K}$ point (see Fig.~\ref{gap_closing}(c)), where the corresponding band energies are relatively high. Hence, the Hall conductivity curve exhibits a comparatively less noticeable deviation near the transition lines. Nevertheless, as the coupling strength $g^l$ increases, the Chern number corresponding to the lower band decreases from positive to negative integer values, indicating the emergence of negative Berry curvature of the lower band, which contributes most significantly to the Hall conductivity. Consequently, the thermal Hall conductivity becomes more negative with increasing coupling strength.

Additionally, in Fig.~\ref{Hall_local}(c), we also illustrate the direct dependence of thermal Hall conductivity ($\kappa_{xy}$) on temperature at three distinct values of $g^l$ (a $g^l =0$ is included for comparison). In the absence of spin-phonon coupling, the Hall conductivity arises purely from magnons. With no spin-phonon coupling, at low temperatures, $\kappa_{xy}$ acquires a positive contribution and reaches a local maximum, beyond which it decreases and eventually becomes negative. It may be noted that, at a finite temperature, $\kappa_{xy}$ vanishes due to the mutual cancellation of contributions from all the bands. Such behavior of the thermal Hall conductivity is well-established in the literature~\cite{Mook2014I, Owerre2017}. Additionally, at very high temperatures, $\kappa_{xy}$ tends to saturate to a finite value.
\begin{figure}[t]
    \centering
    \includegraphics[width=0.95\linewidth]{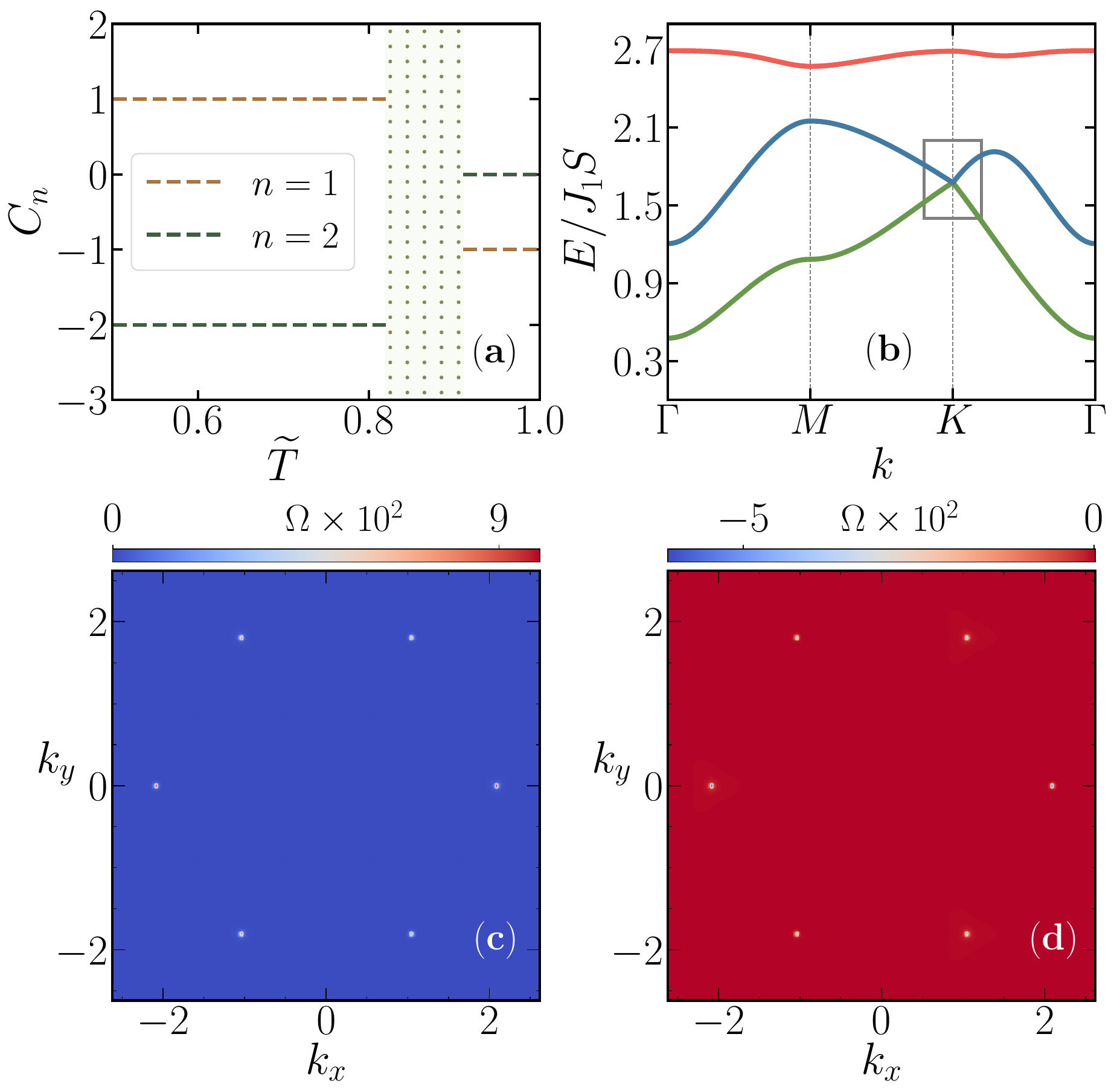}
    \caption{(a) The Chern number ($C_n$) plot corresponding to the lower ($n=1$) and the middle ($n=2$) bands is depicted as a function of $\widetilde{T}$ ($\equiv k_B T/ J_1 S$), where within the dotted region, there occurs closure of gaps between the respective bands. (b) The bulk band structure illustrates the gap closing transition. The Berry curvature plots corresponding to the lower band, (c) before and (d) after the gap closing transition, are shown in the $k_x$-$k_y$ plane.}
    \label{T_transition_local}
\end{figure}

\begin{figure*}[t]
    \centering
    \includegraphics[width=1\linewidth]{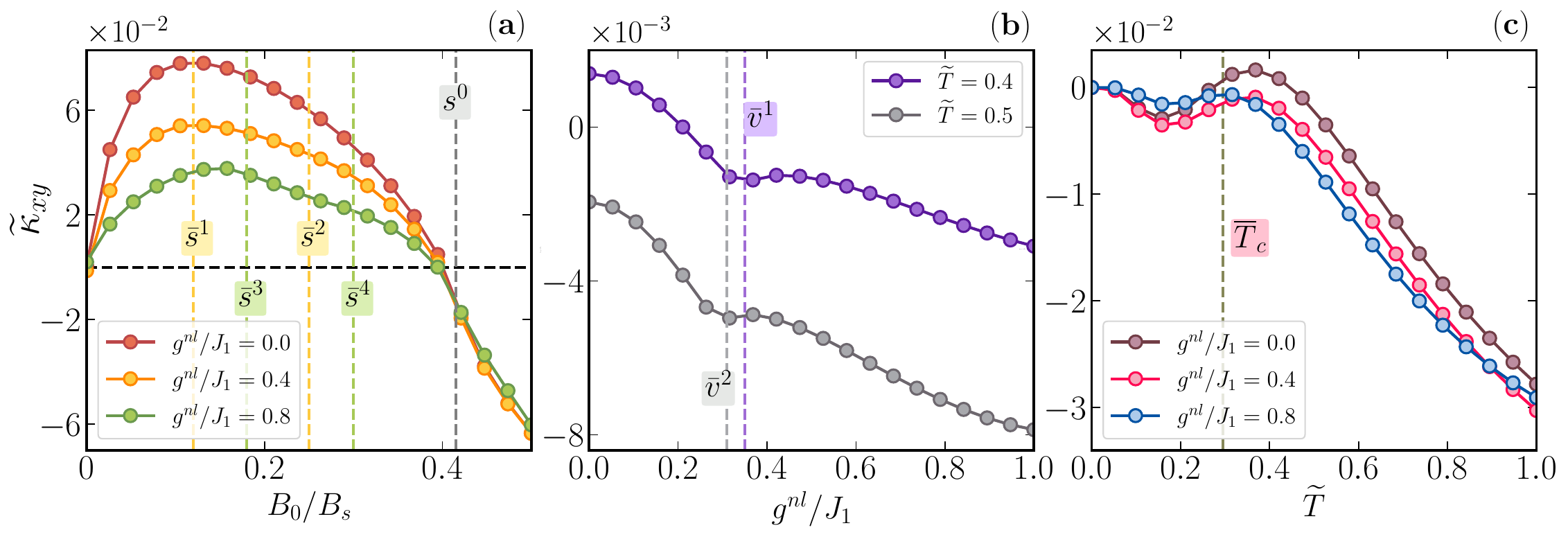}
    \caption{The behavior of the thermal Hall conductivity ($\kappa_{xy}$) in the presence of non-local spin-phonon interaction is demonstrated. (a) $\kappa_{xy}$ is plotted as a function of external magnetic field for different values of the coupling strength $g^{nl}$ at $\widetilde{T}= 0.4$. The transition lines $\bar{s}^1,~\bar{s}^2$ denote $g^{nl}/J_1=0.4$, while $\bar{s}^3,~\bar{s}^4$ refer to $g^{nl}/J_1 = 0.8$, and the $s^0$ transition line corresponds to a topological phase transition for all the cases. (b) $\kappa_{xy}$ is presented as a function of $g^{nl}$ at two different temperatures, keeping $B_0 = 0.4 B_s$, where $\bar{v}^1$ and $\bar{v}^2$ correspond to $\widetilde{T}=0.4$ and $0.8$, respectively. Additionally, (c) shows the thermal Hall conductivity as a function of temperature for distinct values of $g^{nl}$. Here, $\widetilde{\kappa}_{xy} = \kappa_{xy}/J_1S$ and $\widetilde{T} = k_B T/ J_1 S$. Throughout this analysis, the other parameters are fixed at $J_2 = 0.03J_1$, $D = 0.045J_1$, and $\hbar\omega = J_1S$.}
    \label{Hall_nonlocal}
\end{figure*}
In the presence of local spin-phonon coupling ($g^l \neq 0$), the Hall conductivity corresponding to larger values of $g^l$ shows a substantial enhancement at low temperatures, whereas at higher temperatures, the growth becomes comparatively less pronounced as seen for $g^l = 0.8J_1$ compared to $g^l = 0.4J_1$ in Fig.~\ref{Hall_local}(c). While in the absence of spin-phonon coupling, the magnon energy bands are independent of temperature, in its presence, the magnon-polaron states are no longer independent of temperature. As temperature increases, we observe positive energy shifts corresponding to the two lower bands (see Appendix~\ref{finite temperature band structure}), which affects the $c_2(\rho_{n,\mb{k}})$ function in such a way that the overall contribution from $c_2(\rho_{n,\mb{k}})$ function along with temperature responsible for the reduced slope in the behaviour of the thermal Hall conductivity beyond a certain range as it increases. Notably, at higher values of the coupling strength, this behaviour is more prominent. This type of phenomenon reflects the suppression of thermal Hall conductivity due to phonons at high temperatures.

Intriguingly, for $g^l = 0.4 J_1$, we find a topological phase transition from $C = (1, -2, 1)$ to $C = (-1, 0, 1)$ occurring at the transition line $T_c$. For clarity, the corresponding changes in the Chern numbers of the participating bands are presented in Fig.~\ref{T_transition_local}(a) as a function of temperature ($\widetilde{T}$).  Moreover, we observe that the effects of temperature are sufficiently weak such that the gap closing transition at the $\mathbf{K}$ point (see Fig.~\ref{T_transition_local}(b)) extends over a broad temperature range. This region is highlighted by the dotted area in Fig.~\ref{T_transition_local}(a), signifying a weakly semimetalic phase of the bosonic quasiparticles. Owing to the gradual nature of the transition and the fact that it occurs at the $\mathbf{K}$ point, where the contribution from $c_2(\rho_{n, \mathbf{k}})$ is relatively small, the transition is not prominently noticeable from the Hall conductivity plot in Fig.~\ref{Hall_local}(c). Furthermore, in Fig.~\ref{T_transition_local}(c), for the lower band, we show the corresponding Berry curvature prior to the gap closing transition, where positive values of the Berry curvature are concentrated at the high-symmetry $\mb{K}$ and $\mb{K'}$ points, and they become negative beyond the gap closing transition in Fig.~\ref{T_transition_local}(d). This indicates a topological phase transition denoted by $C = 1 \to C = -1$ corresponding to the lower band.   

Similar to the local interaction case, non-local spin-phonon interaction can also lead to some interesting topological phase transitions. The variation of the $\kappa_{xy}$ can be studied as a function of the external magnetic field ($B_0$) for two representative values of $g^{nl}$, while keeping the other parameters fixed at $J_2 = 0.03 J_1$, $D = 0.045 J_1$, $\widetilde{T} = 0.4$. For a direct comparison, we also include the results for the bare magnon case ($g^{l(nl)}=0$), as shown in Fig.~\ref{Hall_nonlocal}(a). As displayed here, at finite $g^{nl}$, namely, $g^{nl} = 0.4J_1$, a gradual increase in $B_0$ leads to a phase transition from $C = (3, -1, -2)$ to $C = (3, -4, 1)$ across the $\bar{s}^1$ transition line at $B_0 = 0.12 B_s$, where the Hall conductivity reaches its peak value. A further increase in $B_0$ induces another phase transition at $\bar{s}^2$ at $B_0 = 0.25B_s$, 
where the Hall conductivity monotonically decreases. A similar sequence of phase transitions is observed for $g^{nl} = 0.8J_1$, as shown in Fig.~\ref{Hall_nonlocal}(a), where the transition lines $\bar{s}^3$ and $\bar{s}^4$ are obtained for  $B_0 = 0.18 B_s$ and $B_0 = 0.3 B_s$, respectively. Moreover,  in contrast to the earlier scenario involving local spin-phonon interactions, where the Hall conductivity for different values of $g^l$ changes sign (from positive to negative) at distinct value of $B_0$, the case corresponding to the non-local interaction exhibits nearly a constant $B_0$ at which the Hall conductivity vanishes for $g^{nl}=0.4J_1$ and $0.8J_1$ (see the clustering of the points on the horizontal dashed line). Consequently, near the $s^0$ transition line, another phase transition from $C = (1, -2, 1)$ to $C = (-1, 0, 1)$ occurs for the cases with $g^{nl} = 0.4J_1$ and $0.8J_1$, as shown in Fig.~\ref{Hall_nonlocal}(a).

It is evident in Fig.~\ref{Phase_plot_nonlocal}(a) that at $\widetilde{T}=0$, with the same values of $D,~J_2$, and $B_0$ mentioned earlier, one phase transition occurs from $C = (3, -4, 1)$ to $C = (1, -2, 1)$ as a function of $g^{nl}$ near $g^{nl} = 0.43J_1$. Hence, in Fig.~\ref{Hall_nonlocal}(b), we show the corresponding nature of the thermal Hall conductivity as a function of $g^{nl}$ at two different temperatures, where we observe a slight non-monotonic behaviour of the Hall conductivity associated with that phase transition across the transition line $\bar{v}^1$ (at $g^{nl}=0.35 J_1$) and $\bar{v}^2$ (at $g^{nl}=0.31 J_1$) corresponding to the temperatures $\widetilde{T} = 0.4$ and $0.5$, respectively. Moreover, it reflects how the transition points shift with varying temperature. Similar to the local case, the Hall conductivity increases with the coupling strength $g^{nl}$ except in the vicinity of the phase transition, as shown in Fig.~\ref{Hall_nonlocal}(b).

Additionally, we inspect the behaviour of Hall conductivity as a function of temperature. Here we keep $B_0 = 0.4 B_s$ same as in Fig.~\ref{Hall_local}(c), and if we see Fig.~\ref{Hall_nonlocal}(a), we can observe near $B_0=0.4 B_s$, the behaviour of the Hall conductivity remains nearly same for all the cases corresponding to different $g^{nl}$ which is manifested Fig.~\ref{Hall_nonlocal}(c). Moreover, there are a couple of crossings between the Hall conductivity curves corresponding to $g^{nl} = 0.4J_1$ (in pink) and $0.8J_1$ (in blue) as seen in Fig.~\ref{Hall_nonlocal}(c). The first crossing at low temperature (near $\widetilde{T}\sim 0.3$) occurs due to a phase transition from $C = (3, -4, 1)$ to $C = (1, -2, 1)$, corresponding to the case with $g^{nl}=0.4J_1$, near the transition line $\overline{T}_c$ at $\widetilde{T} = 0.29$. A second crossing exists at higher values of $\widetilde{T}$, which however is due to the change in the slope of $\kappa_{xy}$ corresponding to the case with $g^{nl}=0.8J_1$, which was earlier denoted as the suppression of the thermal Hall conductivity at high temperature induced by phonons. 

\begin{figure}[t]
    \centering
    \includegraphics[width=0.95\linewidth]{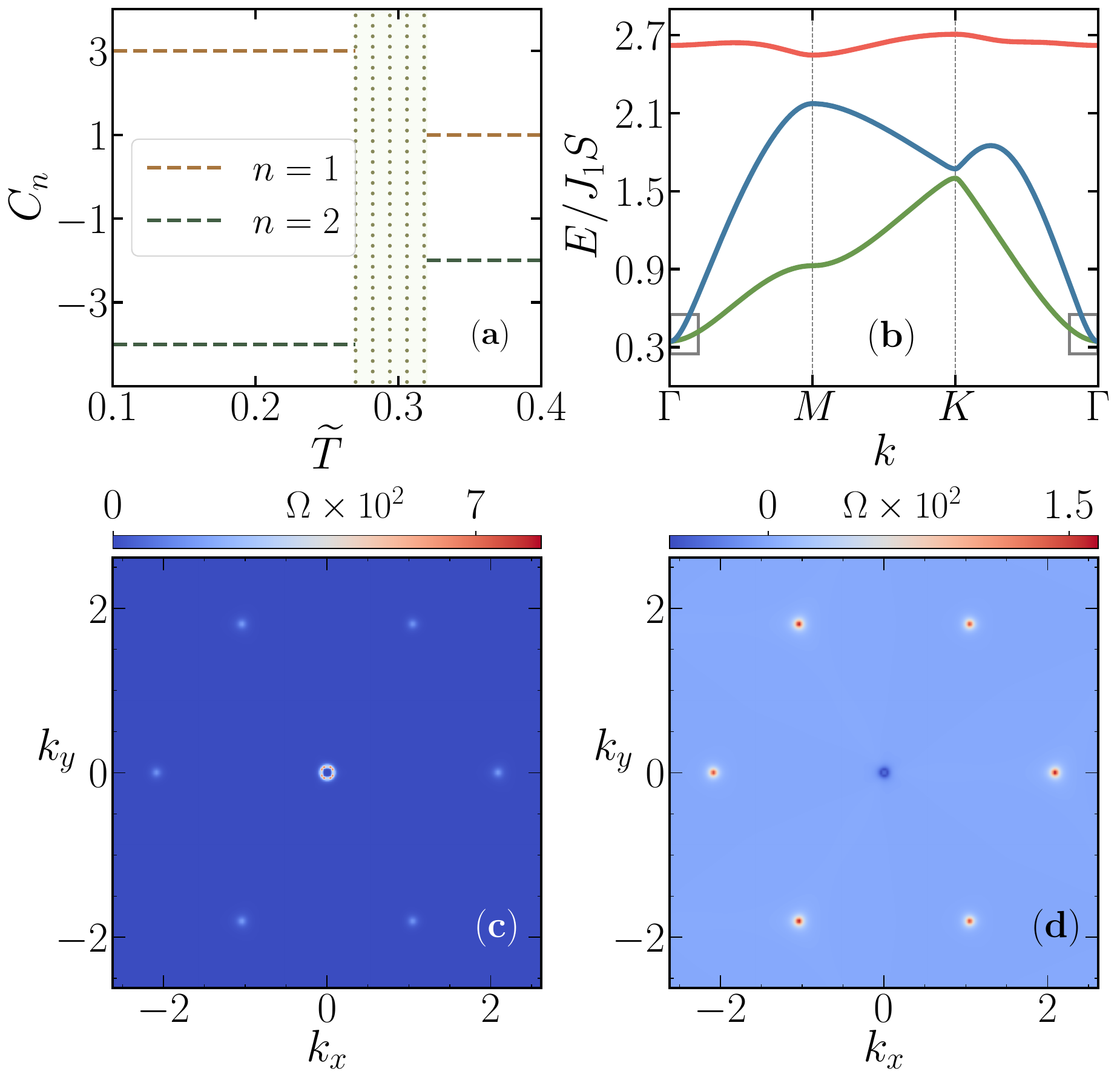}
    \caption{(a) The Chern number ($C_n$) plot corresponding to the lower ($n=1$) and the middle ($n=2$) bands is displayed as a function of $\widetilde{T}$ ($\equiv k_B T/ J_1 S$), where within the dotted region, there occurs closure of gaps between the respective bands, which is further illustrated in (b) via the gap closing transition.
    The Berry curvature plots corresponding to the lower band, (c) before and (d) after the gap closing transition, are shown in the $k_x$-$k_y$ plane.}
    \label{T_transition_nonlocal}
\end{figure}
To clarify the above mentioned phase transition for $g^{nl}=0.4J_1$, we further present the corresponding Chern number plot as a function of temperature in Fig.~\ref{T_transition_nonlocal} (a), where the dotted region corresponds to the spectral gap closing regime. Further, in Fig.~\ref{T_transition_nonlocal}(b), we observe a gap closing transition near the $\mb{\Gamma}$ point, where the contribution from $c_2(\rho_{n, \mb{k}})$ is significant at low temperatures. As a result, the transition appears more prominently in the Hall conductivity plot compared to the case with local spin-phonon interaction. Additionally, in Fig.~\ref{T_transition_nonlocal}(c), at $\widetilde{T}=0.24$, prior to the gap closing transition, we show the Berry curvature of the lower band. Positive values of the Berry curvature are observed near both the high-symmetry points ($\mb{\Gamma}$ and $\mb{K}$), corresponding to a Chern number of $C = 3$. In contrast, in Fig.~\ref{T_transition_nonlocal}(d), after closing the spectral gap, the Berry curvature changes sign  (negative) at the gap closing point $\mb{\Gamma}$, while remaining positive at the $\mb{K}$ point, indicating $C = 1$.

\section{CONCLUSION}
\label{conclusion}
We investigate the effects of couplings between spins and optical phonons in a topologically protected magnetic system, namely, in a frustrated Kagome antiferromagnet that attains a canted configuration under the application of an external magnetic field. Specifically, we consider two different types of spin-phonon couplings, one in which the interaction is purely local, while the other accounts for non-local interactions, to encounter and compare different kinds of topological scenarios. To deal with the LO phonons, we perform detailed and systematic calculations based on a canonical spin-Peierls transformation formulated within the magnon framework and obtain a hybridized magnon-polaron state. Interestingly, we observe several topological phase transitions solely induced by the spin-phonon couplings, which are explored via the magnon-polaron band structures, their Chern numbers, and the corresponding phase plots. Furthermore, the thermal Hall conductivities relevant to these scenarios are obtained, thereby providing robust support for these topological transitions. 

To enumerate the highlights of our results, we observe multiple topological phases accompanied by several bulk gap closing transitions as a function of the DMI and the spin-phonon coupling strength, albeit demonstrating distinct signatures for the local and non-local spin-phonon couplings. To establish the bulk-boundary correspondence, we show the variations of the associated edge modes on a ribbon geometry, where these distinct phases are validated via computing the winding number, which dictates the number of edge crossings and the chirality of the magnon-polaron edge currents. We emphasize that for an identical range of the coupling strengths, the local coupling mechanism renders a larger number of distinct topological phases, compared to the non-local one. Furthermore, in this study, we observe that due to the inclusion of the phonon contributions at finite temperatures, the spectral properties of the energy bands depend on temperature, resulting in topological phase transitions mediated via thermal effects. Moreover, at finite temperatures, the behaviour of the thermal Hall conductivity as a function of the external magnetic field, spin-phonon coupling strength, and temperature, effectively captures all the topological phases and phase transitions occurring therein. The contrasting effects resulting from the two types of couplings further corroborate that local coupling exhibits a comparatively stronger effect on enhancing the thermal Hall conductivity than that arising out of the non-local coupling (particularly when the Hall conductivity nearly vanishes for the bare magnon case). Despite having distinct origins and effects on the topology of the magnon-polaron bands, both types of spin-phonon coupling play crucial roles in shaping the topological features of a frustrated kagome antiferromagnet.

At a fundamental level, our observations give insights into the optical phonon contribution in a topologically protected frustrated magnetic system. While our analyses are based on a specific model, the potential for experimental realization of such a system strengthens the significance of this study.

\section*{Acknowledgments}
S.D. acknowledges financial support from the Ministry of Education (MOE), Govt. of
India. K.B. acknowledges financial support from the Anusandhan National Research Foundation (ANRF), Govt. of India, through the National Post Doctoral Fellowship (NPDF) (File No. PDF/2023/000161).

\appendix 
\section{Renormalized magnon-polaron Hamiltonian from Spin-Phonon Coupling}
\label{Appendix: Magnon polaron Hamiltonian with spin-phonon coupling}
In the following discussion, we incorporate the effect of spin–phonon coupling, which modifies the hopping amplitudes in the presence of a phonon cloud.

\subsection{Decoupling magnon-phonon modes by canonical transformation}
\label{effective Spin-phonon coupling}
As mentioned, we consider two distinct types of spin-phonon interactions, where the local magnons interact with both the local and non-local optical phonon modes. To incorporate the corresponding effects in the effective magnon Hamiltonian, we employ a canonical spin-Peierls transformation~\cite{Weisse1999}, via the operator $U^{l(nl)}=e^{R^{l(nl)}}$, where the transformed Hamiltonian reads as
\begin{equation}
    \widetilde{H}= e^{R^{l(nl)}}He^{-R^{l(nl)}},\label{LFT}
\end{equation}
and the corresponding hybrid wavefunction of the system can be represented as a tensor product of magnonic and phononic wavefunctions, which looks like
\begin{equation}
    |\phi\rangle_\text{hyb}=U^{l(nl)}|\phi\rangle_\text{ph}\otimes|\phi\rangle_\text{mag}.\label{hybwave}
\end{equation}
The unitary generator of the transformation, namely $R^{l(nl)}$ corresponding to the local (non-local) spin-phonon interaction, can be expressed as
\begin{align}
        R^l&= \frac{g^{l}}{\hbar \omega} \sum_{\langle i,j \rangle} \left( \tilde{b}_i^\dagger - \tilde{b}_i \right) \mathbf{S}_i \cdot \mathbf{S}_j,\nonumber\\
        R^{nl}&= \frac{g^{nl}}{\hbar \omega} \sum_{\langle i,j \rangle}\left( \tilde{\beta}_{i,j}^\dagger - \tilde{\beta}_{i,j} \right)\mathbf{S}_i \cdot \mathbf{S}_j.\label{R}
\end{align}
Here, $\tilde{\beta}_{i,j}=\tilde{b}_i-\tilde{b}_j$ and the explicit forms of $R^{l(nl)}$ can be obtained from the HP expansion of $\mathbf{S}_i \cdot \mathbf{S}_j$. Moreover, Eq.~\eqref{LFT} can be cast as a series sum and is given by
\begin{equation}
    \widetilde{H}=H + [R^{l(nl)}, H] + \frac{1}{2!}[R^{l(nl)},[R^{l(nl)}, H]] + \cdots. \label{explicit_form}
\end{equation}
\begin{figure}
    \centering
    \includegraphics[width=0.9\columnwidth]{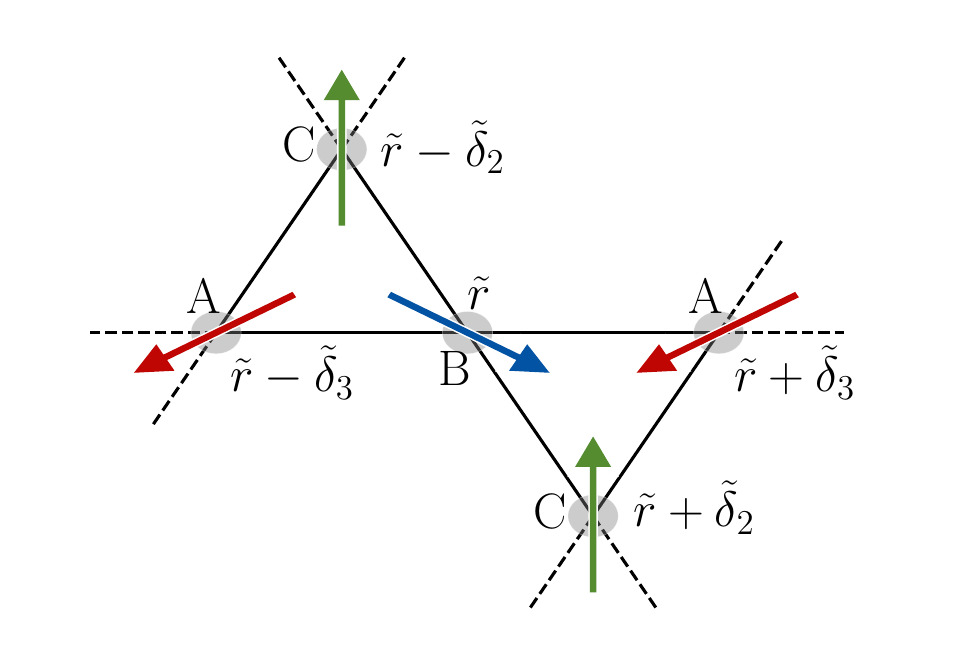}
    \caption{A schematic representation of a generic single unit cell of a KAFM lattice is shown, where each lattice site is labeled with its position coordinate, namely $\tilde{r}\mp\tilde{\delta}_3$, $\tilde{r}$, and $\tilde{r}\mp\tilde{\delta}_2$ for A, B, and C sublattices, respectively with $\tilde{\delta}_i$ ($i\in 1,2,3$) denoting three NN lattice vectors. The circular spots in grey represent the phonon modes corresponding to each site.}
    \label{Kagome_1}
\end{figure}
\noindent Here, $R^{l(nl)}$ satisfies the relation $[R^{l(nl)}, H_\text{ph}]= -H_\text{sp}^{l(nl)}$. Further, it is pertinent to note that in the regime $\hbar\omega \ll J_1S$, that is at very low frequencies, the factor $e^{-R^{l(nl)}}$ vanishes. Consequently, the canonical transformation becomes inapplicable in this limit.

Further, this types of spin-phonon couplings give rise to various types of magnon phonon interaction terms, such as, $\tilde{b}_i^{(\dg)}m_i^\dg m_i$, $\tilde{b}_i^{(\dg)}m_i^\dg m_j$, $\tilde{b}_i^{(\dg)}m_i^\dg m_j^\dg$ ($m \in a, b, c$ corresponding to the three sublattices), etc. Among these, while the onsite coupling terms provide the dominant contribution, the offsite terms lead to scattering processes whose contributions cannot be captured within a well-defined perturbative expansion using non-interacting formalisms. Therefore, as long as we are working within a non-interacting framework, the off-site contributions, both from the spin-phonon interaction Hamiltonian and the corresponding decoupling operators $R^{l(nl)}$, can be safely neglected. Further, we shall show the final form of the decoupling operators below.

In the presence of phonon operators, different lattice points in a unit cell should be marked with different position indices. Hence, in Fig.~(\ref{Kagome_1}), we introduce new dummy index $\tilde{r}$ and the NN lattice vectors, $\tilde{\delta}_i =\frac{\delta_i}{2}$ ($i\in 1,2,3$). Further, the generators in terms of the magnon and phonon bases can be written as
\begin{widetext}
\begin{subequations}
        \begin{align}
            R^l = \alpha^l \sum_{\tilde{r}}&\left[2(\tilde{b}_{\tilde{r}-\tilde{\delta}_3}^\dg - \tilde{b}_{\tilde{r}-\tilde{\delta}_3}) + (\tilde{b}_{\tilde{r}-\tilde{\delta}_2}^\dg - \tilde{b}_{\tilde{r}-\tilde{\delta}_2}) + (\tilde{b}_{\tilde{r}+\tilde{\delta}_2 - 2\tilde{\delta}_3}^\dg - \tilde{b}_{\tilde{r}+\tilde{\delta}_2 - 2\tilde{\delta}_3})\right] a_{\tilde{r}-\tilde{\delta}_3}^\dg a_{\tilde{r}-\tilde{\delta}_3}\nonumber \\
            +&\left[2(\tilde{b}_{\tilde{r}}^\dg - \tilde{b}_{\tilde{r}}) + (\tilde{b}_{\tilde{r}-\tilde{\delta}_3}^\dg - \tilde{b}_{\tilde{r}-\tilde{\delta}_3}) + (\tilde{b}_{\tilde{r}+\tilde{\delta}_3}^\dg - \tilde{b}_{\tilde{r}+\tilde{\delta}_3})\right]b_{\tilde{r}}^\dg b_{\tilde{r}}\nonumber \\
            +&\left[ 2(\tilde{b}_{\tilde{r}-\tilde{\delta}_2}^\dg - \tilde{b}_{\tilde{r}-\tilde{\delta}_2}) 
            + (\tilde{b}_{\tilde{r}-2\tilde{\delta}_2}^\dg - \tilde{b}_{\tilde{r}-2\tilde{\delta}_2}) 
            + (\tilde{b}_{\tilde{r}}^\dg - \tilde{b}_{\tilde{r}}) \right] 
            c_{\tilde{r}-\tilde{\delta}_2}^\dg c_{\tilde{r}-\tilde{\delta}_2},
        \end{align}
    \begin{align}
        R^{nl} = \alpha^{nl} \sum_{\tilde{r}}&\left[(\tilde{b}_{\tilde{r}-\tilde{\delta}_2}^\dg - \tilde{b}_{\tilde{r}-\tilde{\delta}_2}) - (\tilde{b}_{\tilde{r}}^\dg - \tilde{b}_{\tilde{r}}) + (\tilde{b}_{\tilde{r}+\tilde{\delta}_2 - 2\tilde{\delta}_3}^\dg - \tilde{b}_{\tilde{r}+\tilde{\delta}_2 - 2\tilde{\delta}_3}) - (\tilde{b}_{\tilde{r}-2\tilde{\delta}_3}^\dg - \tilde{b}_{\tilde{r}-2\tilde{\delta}_3})\right]a_{\tilde{r}-\tilde{\delta}_3}^\dg a_{\tilde{r}-\tilde{\delta}_3}\nonumber \\
        + &\left[(\tilde{b}_{\tilde{r}-\tilde{\delta}_3}^\dg - \tilde{b}_{\tilde{r}-\tilde{\delta}_3}) - (\tilde{b}_{\tilde{r}-\tilde{\delta}_2}^\dg - \tilde{b}_{\tilde{r}-\tilde{\delta}_2}) + (\tilde{b}_{\tilde{r}+\tilde{\delta}_3}^\dg - \tilde{b}_{\tilde{r}+\tilde{\delta}_3}) - (\tilde{b}_{\tilde{r}+\tilde{\delta}_2}^\dg - \tilde{b}_{\tilde{r}+\tilde{\delta}_2})\right]b_{\tilde{r}}^\dg b_{\tilde{r}}\nonumber \\
        + & \left[(\tilde{b}_{\tilde{r}}^\dg - \tilde{b}_{\tilde{r}}) - (\tilde{b}_{\tilde{r}-\tilde{\delta}_3}^\dg - \tilde{b}_{\tilde{r}-\tilde{\delta}_3}) + (\tilde{b}_{\tilde{r}-2\tilde{\delta}_2}^\dg - \tilde{b}_{\tilde{r}-2\tilde{\delta}_2}) - (\tilde{b}_{\tilde{r}-2\tilde{\delta}_2 + \tilde{\delta}_3}^\dg - \tilde{b}_{\tilde{r}-2\tilde{\delta}_2 + \tilde{\delta}_3})\right]c_{\tilde{r}-\tilde{\delta}_2}^\dg c_{\tilde{r}-\tilde{\delta}_2},
    \end{align}
\end{subequations}
where $\alpha^{l(nl)} = (g^{l(nl)}/\hbar \omega) S\mathcal{E}$. Incorporating this into Eq.~\eqref{explicit_form}, we obtain the generalized form of the modified Hamiltonian \( \widetilde{H} \) in terms of the general magnon basis \( m \) (\( m \in \{a, b, c\} \)), which is given by
\begingroup
\allowdisplaybreaks
\begin{equation}
\begin{split}
\widetilde{H} = S &\left[ 
     \sum_i \left( 4(J_1 + J_2)\,{\mathcal{E}} + 4D\,{\mathcal{E}}_1 + \mu_B B_0 \sin{\eta} - \Delta^{l(nl)} \right) m_i^\dagger m_i 
    + \hbar \omega \sum_i \tilde{b}_i^\dagger \tilde{b}_i \right. \\
     &\left. +\sum_{\langle ij \rangle} \left( J_1 (P + iQ) - D (P_1 + iQ_1) \right) m_i^\dagger m_j e^{X_1^{l(nl)}}
    + \left( J_1 R - D R_1 \right) m_i^\dagger m_j^\dagger e^{X_2^{l(nl)}} \right. \\
     &\left. +\sum_{\langle\langle ij \rangle\rangle} \left( J_2 (P + iQ) \right) m_i^\dagger m_j e^{X_3^{l(nl)}}
    + J_2 R m_i^\dagger m_j^\dagger e^{X_4^{l(nl)}},
\right]
\end{split}\label{tilde_H_LFT}
\end{equation}
\endgroup
\end{widetext}
\noindent where $n$ and $nl$ correspond to the local and non-local spin-phonon interactions, respectively. While 
\begin{equation}
    \Delta^l = 6 (\alpha^{l})^2 \hbar \omega,~\Delta^{nl} = 4 (\alpha^{nl})^2 \hbar \omega
\end{equation}
are independent of the position in the lattice, the terms $X_q^{l(nl)}$ ($q\in 1,2 ,3 ,4$) appearing in the exponents depend explicitly on the positions of the magnons. For clarity and conciseness, we present the explicit forms of $X_q^{l(nl)}$ for a representative bond within a single unit cell. To demonstrate this, we consider \( a_{\tilde{r}-\tilde{\delta}_3}^\dagger b_{\tilde{r}}e^{X_1^{l(nl)}} \) and \( a_{\tilde{r}-\tilde{\delta}_3}^\dagger b_{\tilde{r}}^\dagger e^{X_2^{l(nl)}} \) as representative NN interactions, and \( a_{\tilde{r}-\tilde{\delta}_3}^\dagger b_{\tilde{r}-2\tilde{\delta}_2} e^{X_3^{l(nl)}}\) and \( a_{\tilde{r}-\tilde{\delta}_3}^\dagger b_{\tilde{r}-2\tilde{\delta}_2}^\dagger e^{X_4^{l(nl)}}\) as NNN interactions. For the local interaction, the corresponding terms $X_q^l$ can be written as
\begin{widetext}
    \begingroup
    \allowdisplaybreaks
    \begin{align}
        X_1^l &= \alpha^{l} \Big[
        \left(\tilde{b}_{\tilde{r}-\tilde{\delta}_3}^\dagger - \tilde{b}_{\tilde{r}-\tilde{\delta}_3}\right) 
        + \left(\tilde{b}_{\tilde{r}-\tilde{\delta}_2}^\dagger - \tilde{b}_{\tilde{r}-\tilde{\delta}_2}\right) 
        + \left(\tilde{b}_{\tilde{r}+\tilde{\delta}_2 - 2\tilde{\delta}_3}^\dagger - \tilde{b}_{\tilde{r}+\tilde{\delta}_2 - 2\tilde{\delta}_3}\right) - 2\left(\tilde{b}_{\tilde{r}}^\dagger - \tilde{b}_{\tilde{r}}\right) 
        - \left(\tilde{b}_{\tilde{r}+\tilde{\delta}_3}^\dagger - \tilde{b}_{\tilde{r}+\tilde{\delta}_3}\right)
    \Big],  \nonumber\\
         X_2^l &= \alpha^{l} \Big[
        3\left(\tilde{b}_{\tilde{r}-\tilde{\delta}_3}^\dagger - \tilde{b}_{\tilde{r}-\tilde{\delta}_3}\right) 
        + \left(\tilde{b}_{\tilde{r}-\tilde{\delta}_2}^\dagger - \tilde{b}_{\tilde{r}-\tilde{\delta}_2}\right) 
        + \left(\tilde{b}_{\tilde{r}+\tilde{\delta}_2 - 2\tilde{\delta}_3}^\dagger - \tilde{b}_{\tilde{r}+\tilde{\delta}_2 - 2\tilde{\delta}_3}\right) + 2\left(\tilde{b}_{\tilde{r}}^\dagger - \tilde{b}_{\tilde{r}}\right) 
        + \left(\tilde{b}_{\tilde{r}+\tilde{\delta}_3}^\dagger - \tilde{b}_{\tilde{r}+\tilde{\delta}_3}\right)
    \Big],  \nonumber\\
        X_3^l &= \alpha^{l} \Big[
        2\left(\tilde{b}_{\tilde{r}-\tilde{\delta}_3}^\dagger - \tilde{b}_{\tilde{r}-\tilde{\delta}_3}\right)
        + \left(\tilde{b}_{\tilde{r}-\tilde{\delta}_2}^\dagger - \tilde{b}_{\tilde{r}-\tilde{\delta}_2}\right) 
        + \left(\tilde{b}_{\tilde{r}+\tilde{\delta}_2 - 2\tilde{\delta}_3}^\dagger - \tilde{b}_{\tilde{r}+\tilde{\delta}_2 - 2\tilde{\delta}_3}\right) - 2 \left(\tilde{b}_{\tilde{r}-2\tilde{\delta}_2}^\dagger - \tilde{b}_{\tilde{r}-2\tilde{\delta}_2}\right)\nonumber \\
        &\hspace{7cm}- \left(\tilde{b}_{\tilde{r}-2\tilde{\delta}_2 - \tilde{\delta}_3}^\dagger - \tilde{b}_{\tilde{r}-2\tilde{\delta}_2 - \tilde{\delta}_3}\right) - \left(\tilde{b}_{\tilde{r}-2\tilde{\delta}_2 + \tilde{\delta}_3}^\dagger - \tilde{b}_{\tilde{r}-2\tilde{\delta}_2 + \tilde{\delta}_3}\right)
    \Big], \nonumber\\
       X_4^l &= \alpha^{l} \Big[
        2\left(\tilde{b}_{\tilde{r}-\tilde{\delta}_3}^\dagger - \tilde{b}_{\tilde{r}-\tilde{\delta}_3}\right)
        + \left(\tilde{b}_{\tilde{r}-\tilde{\delta}_2}^\dagger - \tilde{b}_{\tilde{r}-\tilde{\delta}_2}\right) 
        + \left(\tilde{b}_{\tilde{r}+\tilde{\delta}_2 - 2\tilde{\delta}_3}^\dagger - \tilde{b}_{\tilde{r}+\tilde{\delta}_2 - 2\tilde{\delta}_3}\right) + 2 \left(\tilde{b}_{\tilde{r}-2\tilde{\delta}_2}^\dagger - \tilde{b}_{\tilde{r}-2\tilde{\delta}_2}\right)\nonumber \\
        &\hspace{7cm}+ \left(\tilde{b}_{\tilde{r}-2\tilde{\delta}_2 - \tilde{\delta}_3}^\dagger - \tilde{b}_{\tilde{r}-2\tilde{\delta}_2 - \tilde{\delta}_3}\right) + \left(\tilde{b}_{\tilde{r}-2\tilde{\delta}_2 + \tilde{\delta}_3}^\dagger - \tilde{b}_{\tilde{r}-2\tilde{\delta}_2 + \tilde{\delta}_3}\right)
    \Big],
    \end{align}
whereas, for the non-local interaction, $X_q^{nl}$ can be written as,
\begin{align}
    X_1^{nl} &= \alpha^{nl} \Big[
        2\left(\tilde{b}_{\tilde{r}-\tilde{\delta}_2}^\dagger - \tilde{b}_{\tilde{r}-\tilde{\delta}_2}\right)
        + \left(\tilde{b}_{\tilde{r}+\tilde{\delta}_3}^\dagger - \tilde{b}_{\tilde{r}+\tilde{\delta}_3}\right)
        +\left(\tilde{b}_{\tilde{r}+\tilde{\delta}_2 - 2\tilde{\delta}_3}^\dagger - \tilde{b}_{\tilde{r}+\tilde{\delta}_2 - 2\tilde{\delta}_3}\right)
        - \left(\tilde{b}_{\tilde{r}}^\dagger - \tilde{b}_{\tilde{r}}\right)\nonumber \\
        &\hspace{6cm}- \left(\tilde{b}_{\tilde{r}-2\tilde{\delta}_3}^\dagger - \tilde{b}_{\tilde{r}-2\tilde{\delta}_3}\right)
        -\left(\tilde{b}_{\tilde{r}-\tilde{\delta}_3}^\dagger - \tilde{b}_{\tilde{r}-\tilde{\delta}_3}\right)
        -\left(\tilde{b}_{\tilde{r}+\tilde{\delta}_2}^\dagger - \tilde{b}_{\tilde{r}+\tilde{\delta}_2}\right)
    \Big],\nonumber \\
    X_2^{nl} &= \alpha^{nl} \Big[
        \left(\tilde{b}_{\tilde{r}-\tilde{\delta}_3}^\dagger - \tilde{b}_{\tilde{r}-\tilde{\delta}_3}\right)
        + \left(\tilde{b}_{\tilde{r}+\tilde{\delta}_3}^\dagger - \tilde{b}_{\tilde{r}+\tilde{\delta}_3}\right)
        + \left(\tilde{b}_{\tilde{r}+\tilde{\delta}_2 - 2\tilde{\delta}_3}^\dagger - \tilde{b}_{\tilde{r}+\tilde{\delta}_2 - 2\tilde{\delta}_3}\right)
        - \left(\tilde{b}_{\tilde{r}}^\dagger - \tilde{b}_{\tilde{r}}\right)\nonumber \\
        &\hspace{6cm}- \left(\tilde{b}_{\tilde{r}-2\tilde{\delta}_3}^\dagger - \tilde{b}_{\tilde{r}-2\tilde{\delta}_3}\right)
        - \left(\tilde{b}_{\tilde{r}+\tilde{\delta}_2}^\dagger - \tilde{b}_{\tilde{r}+\tilde{\delta}_2}\right)
    \Big],\nonumber \\
    X_3^{nl} &= \alpha^{nl} \Big[
        2\left(\tilde{b}_{\tilde{r}-\tilde{\delta}_2}^\dagger - \tilde{b}_{\tilde{r}-\tilde{\delta}_2}\right)
        + \left(\tilde{b}_{\tilde{r}+\tilde{\delta}_2 - 2\tilde{\delta}_3}^\dagger - \tilde{b}_{\tilde{r}+\tilde{\delta}_2 - 2\tilde{\delta}_3}\right)
        + \left(\tilde{b}_{\tilde{r}-3\tilde{\delta}_2}^\dagger - \tilde{b}_{\tilde{r}-3\tilde{\delta}_2}\right)
        - \left(\tilde{b}_{\tilde{r}}^\dagger - \tilde{b}_{\tilde{r}}\right)
        - \left(\tilde{b}_{\tilde{r}-2\tilde{\delta}_3}^\dagger - \tilde{b}_{\tilde{r}-2\tilde{\delta}_3}\right)\nonumber \\
        &\hspace{6cm} - \left(\tilde{b}_{\tilde{r}-2\tilde{\delta}_2 - \tilde{\delta}_3}^\dagger - \tilde{b}_{\tilde{r}-2\tilde{\delta}_2 - \tilde{\delta}_3}\right)
        - \left(\tilde{b}_{\tilde{r}-2\tilde{\delta}_2 + \tilde{\delta}_3}^\dagger - \tilde{b}_{\tilde{r}-2\tilde{\delta}_2 + \tilde{\delta}_3}\right)
    \Big],\nonumber \\
    X_4^{nl} &= \alpha^{nl} \Big[
        \left(\tilde{b}_{\tilde{r}+\tilde{\delta}_2 - 2\tilde{\delta}_3}^\dagger - \tilde{b}_{\tilde{r}+\tilde{\delta}_2 - 2\tilde{\delta}_3}\right)
        +\left(\tilde{b}_{\tilde{r}-2\tilde{\delta}_2 - \tilde{\delta}_3}^\dagger - \tilde{b}_{\tilde{r}-2\tilde{\delta}_2 - \tilde{\delta}_3}\right)
        +\left(\tilde{b}_{\tilde{r}-2\tilde{\delta}_2 + \tilde{\delta}_3}^\dagger - \tilde{b}_{\tilde{r}-2\tilde{\delta}_2 + \tilde{\delta}_3}\right)
        - \left(\tilde{b}_{\tilde{r}}^\dagger - \tilde{b}_{\tilde{r}}\right)\nonumber \\
        &\hspace{6cm} -\left(\tilde{b}_{\tilde{r}-2\tilde{\delta}_3}^\dagger - \tilde{b}_{\tilde{r}-2\tilde{\delta}_3}\right)
        - \left(\tilde{b}_{\tilde{r}-3\tilde{\delta}_2}^\dagger - \tilde{b}_{\tilde{r}-3\tilde{\delta}_2}\right)
    \Big].
\end{align}
    \endgroup
Although the terms $X_q^{l(nl)}$ differ for other NN or NNN bonds, the number of phonon degrees of freedom remains the same in each of the cases. Throughout the calculations presented in our work, we have neglected the four-magnon quartic terms such as $m_i^\dagger m_j^\dagger m_i m_j$ ($\forall i,j$). 

\subsection{Reduced Hamiltonian via finite-temperature phonon averaging}
\label{renormalized Hamiltonian}
\begingroup
\allowdisplaybreaks
Now, to eliminate the phonon degrees of freedom, we take a finite-temperature phonon average of the transformed Hamiltonian (Eq.~\eqref{tilde_H_LFT}), namely
\begin{equation}
    \widetilde{H}_\text{R} = \braket{\widetilde{H}}_\mathrm{ph}.\label{H_R}
\end{equation}
Here, $\widetilde{H}$ contains both phonon operators and magnon operators, and only the phonon operators take part in this averaging. To further elaborate on the method, we consider a generic phonon operator of the form
\begin{equation}
    \widetilde{O}= e^{\alpha(\tilde{b}_r^\dg - \tilde{b}_r)},
\end{equation}
where $\tilde{b}_r$ is the phonon operator corresponds to the lattice site $r$, $\alpha$ denotes the coupling strength, and the associated phonon energy of that site is $H_{\mathrm{ph}} = \hbar \omega (\tilde{b}_r^\dg\tilde{b}_r)$. Then, the thermal phonon average of $\widetilde{O}$ with respect to the $n$-th harmonic oscillator state, namely $\ket{n}$ can be expressed as
\begin{equation}
    \braket{\widetilde{O}} = \frac{\sum_{n=0}^\infty \bra{n}e^{-\beta H_{\mathrm{ph}}}\widetilde{O}\ket{n}}{\sum_{n=0}^\infty \bra{n}e^{-\beta H_{\mathrm{ph}}}\ket{n}}= \left[e^{\beta \hbar \omega} f_{\mathrm{ph}}\right]^{-1}\sum_{n=0}^\infty e^{-n\beta \hbar \omega}\bra{n}\widetilde{O}\ket{n}, \label{avg_O}
\end{equation}
where $\beta = 1/k_BT$ and $f_{\mathrm{ph}}= 1/(e^{\beta\hbar \omega}-1)$ is the BE distribution function. Now, the expectation of $\widetilde{O}$ in Eq.~\eqref{avg_O} yields~\cite{Glauber1963,Scully1997}
\begin{equation}
    \bra{n}\widetilde{O}\ket{n} = e^{-\frac{1}{2}\alpha^2}\bra{n}e^{\alpha \tilde{b}_r^\dg}e^{-\alpha \tilde{b}_r}\ket{n},
\end{equation}
where the action of $\tilde{b}_r^m$ on $n$-th harmonic oscillator state reads as

\begin{equation}
    \tilde{b}_r^m \ket{n}=\sqrt{\frac{n!}{(n-m)!}}\ket{n-m}.
\end{equation}
Finally, Eq.~\eqref{avg_O} takes the form
\begin{equation}
    \braket{\widetilde{O}}= \left[e^{\beta \hbar \omega} f_{\mathrm{ph}}\right]^{-1}e^{-\frac{\alpha^2}{2}}\sum_{n=0}^\infty e^{-n\beta \hbar \omega}\mathcal{L}_n(\alpha^2) = e^{-\frac{1}{2}\alpha^2(2f_{\mathrm{ph}} + 1)},
\end{equation}
where $\mathcal{L}_n(\alpha^2)$ is the Laguerre polynomial function. Further, applying the thermal averaging in Eq.~\eqref{H_R}, the reduced Hamiltonian is obtained as
\endgroup
    \begingroup
    \allowdisplaybreaks
    \begin{align}
        \widetilde{H}_\text{R} = S &\left[ 
     \sum_i \left( 4(J_1 + J_2)\,{\mathcal{E}} + 4D\,{\mathcal{E}}_1 + \mu_B B_0 \sin{\eta} - \Delta^{l(nl)} \right) m_i^\dagger m_i 
    \right.\nonumber \\
     &\left. +\sum_{\langle ij \rangle} \left( J_1 (P + iQ) - D (P_1 + iQ_1) \right) m_i^\dagger m_j e^{-\lambda_1^{l(nl)}}
    + \left( J_1 R - D R_1 \right) m_i^\dagger m_j^\dagger e^{-\lambda_2^{l(nl)}} \right.\nonumber \\
     &\left. +\sum_{\langle\langle ij \rangle\rangle} \left( J_2 (P + iQ) \right) m_i^\dagger m_j e^{-\lambda_3^{l(nl)}}
    + J_2 R m_i^\dagger m_j^\dagger e^{-\lambda_4^{l(nl)}}
\right],\label{H_R_eff}
    \end{align}
    where
    \begin{align}
        \lambda_1^l &= 4 (\alpha^l)^2 (2 f_{\text{ph}}+ 1), ~\lambda_2^l = 8 (\alpha^l)^2 (2 f_{\text{ph}}+ 1), ~\lambda_3^l = \lambda_4^l = 6 (\alpha^l)^2 (2 f_{\text{ph}}+ 1),\nonumber\\
        \lambda_1^{nl} &= \lambda_3^{nl} = 5 (\alpha^{nl})^2 (2 f_{\text{ph}}+ 1), \lambda_2^{nl} =\lambda_4^{nl}= 3 (\alpha^{nl})^2 (2 f_{\text{ph}}+ 1).
    \end{align}
The exponential factors, $e^{-\lambda_q^l}$ ($q\in 1,2,3,4$) are often termed as Holstein reduction factors that are widely explored in the context of band-narrowing effects in narrow-band electronic systems~\cite{Bhattacharyya2024,Lu2023}, whereas $\Delta^{l(nl)}$ acts like a polaronic shift energy for the magnons. Thus, the magnon-polaron hybridized bound state is formed, where the hybridized bands are renormalized by the spin-phonon coupling strength.  Moreover, the momentum-space form of the reduced Hamiltonian is provided in Sec.~\ref{Formalism}.
\endgroup
\end{widetext}

\section{Phase diagram for a higher value of $J_2$}
\label{Appendix_Phase_diagrams}
In the main text, we focused on results obtained for small values of $J_2$ and $D$, since a finite $J_2 (> 0)$ along with a moderate out-of-plane DMI can stabilize the $120^\circ$ coplanar spin configuration in KAFM. Additionally, these parameters play a crucial role in determining the emergence of non-coplanar ground state ordering in the presence of an external magnetic field. 

\begin{figure}[H]
    \centering
    \includegraphics[width=1\linewidth]{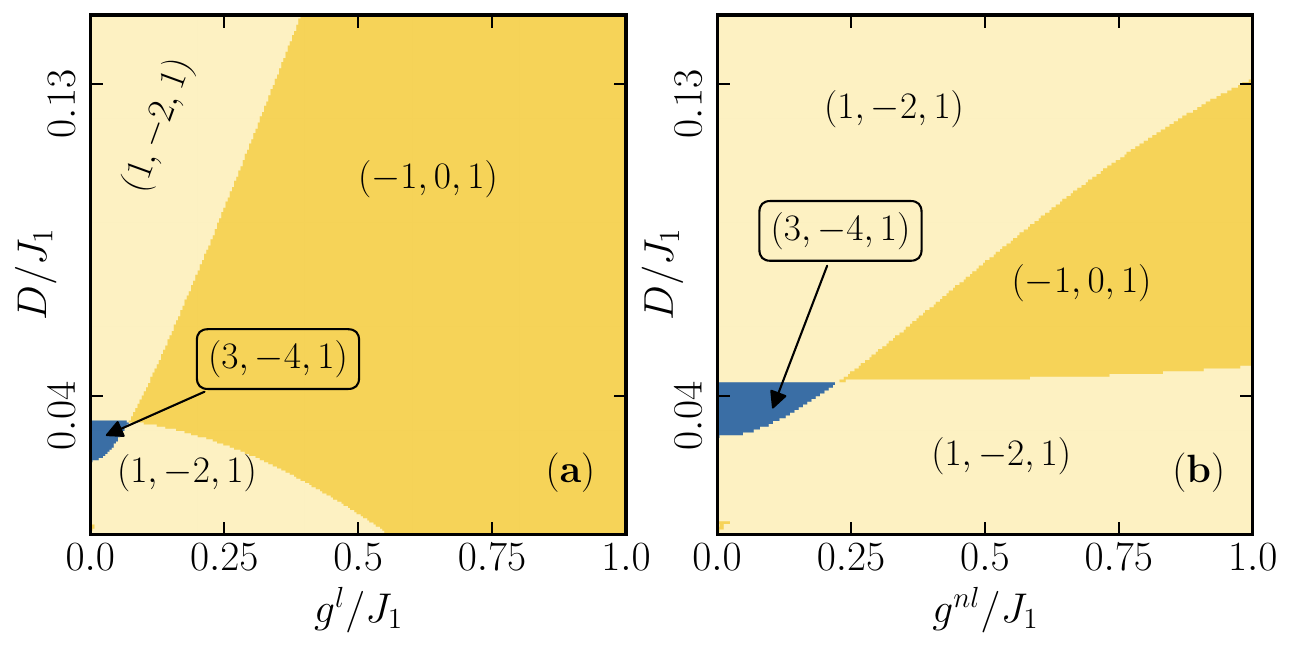}
    \caption{The combined phase diagrams are presented for $J_2 = 0.1J_1$, $B_0= 0.4B_s$, $\widetilde{T}=0$, $\hbar \omega = J_1S$ in the (a) $g^l$-$D$ and (b) $g^{nl}$-$D$ plane, where both $g^{l(nl)}$ and $D$ are measured in units of $J_1$. the Chern number set ($C_1,C_2,C_3$) is designated for each of the topological phases.}
    \label{Additional_Phase_plots}
\end{figure}

Hence, to provide a more comprehensive view of the parameter space, we present the corresponding phase diagrams at a larger coupling strength, $J_2 = 0.1J_1$, as functions of $D$ and $g^l$ in Fig.\ref{Additional_Phase_plots}(a), and $D$ and $g^{nl}$ in Fig.\ref{Additional_Phase_plots}(b). Although in both cases, the phase plots emerge with the same number of distinct topological phases, the phase boundaries differ from each other owing to different coupling mechanisms. Notably, for these higher values of $J_2$ and $D$, the corresponding $\omega_c$ would be higher than the values shown in Fig.~\ref{omega_c_with_g}. Hence, we keep $\hbar \omega = J_1S$ for both the cases.

\section{Band Structures at finite temperature}
\label{finite temperature band structure}
In the main text, for generality, all the band structures are shown in Sec.~\ref{results:bulk} at zero temperature. However, from Eq.~\ref{H_R_kspace}, it is clear that temperature enters into the Hamiltonian via the hopping energies, and consequently it affects the thermal Hall conductivity. Hence, we present additional band structure plots at finite temperatures in the presence of local and non-local spin-phonon couplings.

\begin{figure}[H]
    \centering
    \includegraphics[width=0.95\linewidth]{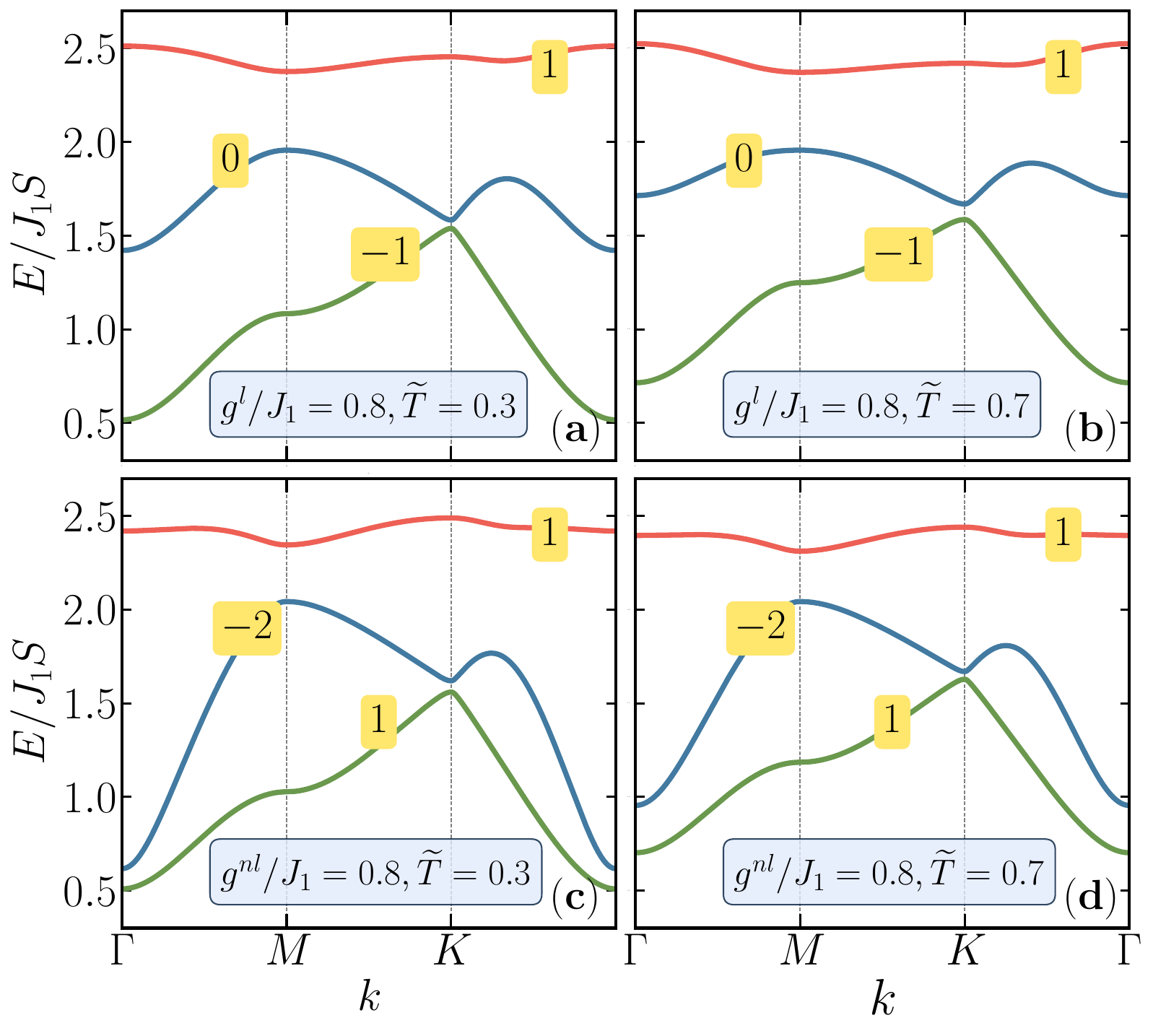}
    \caption{In the presence of local spin-phonon coupling, the band structures (with the corresponding Chern numbers highlighted in yellow) are shown for (a) $g^{l} = 0.8J_1,~\widetilde{T}=0.3$ and (b) $g^{l} = 0.8J_1,~\widetilde{T}=0.7$. Further, in the presence of non-local spin-phonon coupling, the band structures are shown for (c) $g^{nl} = 0.8J_1,~\widetilde{T}=0.3$ and (d) $g^{nl} = 0.8J_1,~\widetilde{T}=0.7$. The other parameters are fixed at $B_0 = 0.4 B_s$, $J_2 = 0.03J_1$, $D = 0.045 J_1$ and $\hbar\omega = J_1 S$. }
    \label{Temperature_bandstructure}
\end{figure}

In the presence of local spin-phonon coupling, we observe that the two lower bands (that is lower and middle) exhibit a positive energy shift and the band gap between them increases near the high symmetry points (see Figs.~\ref{Temperature_bandstructure}(a) and~\ref{Temperature_bandstructure}(b)) as a function of temperature. In contrast, in the presence of the non-local coupling, the band gap near the $\mb{K}$ point decreases, which raises the possibility of another gap closing transition at higher temperatures. However, here also the two lower bands exhibit a positive energy shift as shown in Figs.~\ref{Temperature_bandstructure}(c) and~\ref{Temperature_bandstructure}(d).
Thus, if magnon-polaron instability arises at higher values of the phonon frequency $\omega$, the situation can be stabilized via raising the temperature.

However, since magnon fluctuations, decay processes, and scattering become significant at high temperatures ($k_B T\gtrsim J_1S$), which have been neglected in this study, the stabilization is effective only in the low-temperature regime.
\vspace{4pt}
\bibliography{main}

\end{document}